\documentclass[preprint,trackchanges,longauthor,times]{aastex62}

\hypersetup{linkcolor=red,citecolor=blue,filecolor=cyan,urlcolor=magenta}

\newcommand{\lrsp}{   {Living Reviews in Solar Physics}}

\received{June 9, 2018}
\revised{August 9, 2018}
\accepted{August 17, 2018}

\usepackage{amsmath}

\submitjournal{ApJ}

\shorttitle{Thermodynamic state of a CME}
\shortauthors{Mishra et al.}

\begin{document}

\title{Modeling the thermodynamic evolution of Coronal Mass Ejections (CMEs) using their kinematics}

%\correspondingauthor{Wageesh Mishra}
%\email{wageesh@ustc.edu.cn}

\author[0000-0003-2740-2280]{Wageesh Mishra}
\affil{CAS Key Laboratory of Geospace Environment, Department of Geophysics and Planetary Sciences, \\
University of Science and Technology of China, Hefei 230026, China; \textcolor{blue}{wageesh@ustc.edu.cn}}

\author{Yuming Wang}
\affil{CAS Key Laboratory of Geospace Environment, Department of Geophysics and Planetary Sciences, \\
University of Science and Technology of China, Hefei 230026, China; \textcolor{blue}{ymwang@ustc.edu.cn}}

\begin{abstract}
Earlier studies on Coronal Mass Ejections (CMEs), using remote sensing and in situ observations, have attempted to determine some of the internal properties of CMEs, which were limited to a certain position or a certain time. For understanding the evolution of the internal thermodynamic state of CMEs during their heliospheric propagation, we improve the self-similar flux rope internal state (FRIS) model, which is constrained by measured propagation and expansion speed profiles of a CME. We implement the model to a CME erupted on 2008 December 12 and probe the internal state of the CME. It is found that the polytropic index of the CME plasma decreased continuously from 1.8 to 1.35 as the CME moved away from the Sun, implying that the CME released heat before it reached adiabatic state and then absorbed heat. We further estimate the entropy changing and heating rate of the CME. We also find that the thermal force inside the CME is the internal driver of CME expansion while Lorentz force prevented the CME from expanding. It is noted that centrifugal force due to poloidal motion decreased with the fastest rate and Lorentz force decreased slightly faster than thermal pressure force as CME moved away from the Sun. We also discuss the limitations of the model and approximations made in the study.
\end{abstract}

\textbf{\section{Introduction}\label{intro}} 

Coronal mass ejections (CMEs) are the large-scale structures containing mass, kinetic energy and magnetic flux that are expelled from the Sun into the heliosphere \citep{Tousey1973,Hundhausen1984,Chen2011,Webb2012}. The CMEs in the heliosphere are referred as interplanetary coronal mass ejections (ICMEs) which are often understood as flux-ropes maintaining their magnetic connection to the Sun while propagating in the solar wind \citep{Larson1997}. They are of interest to scientific community because studying them we can enhance our understanding of (i) physical process responsible for removal of built-up magnetic energy and plasma from the solar corona (ii) kinematic and thermodynamic evolution of an expanding magnetized plasma blob in an ambient magneto-fluid medium and (iii) mechanism for exchange of energy and plasma dynamics between different plasma regions \citep{Low2001,Howard2007,Lopez2000}. They are also of interest to the technical community because they are the main drivers of severe space weather leading to disruption to a range of technologies such as communications systems, electronic circuits and power grids at the 
Earth, and thus our technology is vulnerable to them \citep{Schwenn2006,Pulkkinen2007,Baker2009}.

CMEs have been detected by remote sensing and in-situ spacecraft observations for decades. Until a decade back, CMEs were routinely imaged only near the Sun primarily
by coronagraphs onboard spacecraft. The launch of \textit{Solar TErrestrial RElations Observatory} (STEREO) in 2006 \citep{Kaiser2008} provided us the opportunity to track CMEs continuously between the Sun and Earth from multiple viewpoints. Most of the studies have focused on the signatures of origins for CMEs, their dynamic evolutions, masses, arrival times, induced radio bursts, geo-effectiveness, and connection between magnetic flux ropes measured in-situ and CME structures observed by coronagraphs and heliospheric imagers  \citep{Munro1979,Webb2000,Wang2002,Cliver2002,Schwenn2005,Forsyth2006,Gopalswamy2009,Kilpua2012,Deforest2013,Mishra2015b,Shen2017,Mishra2017,Harrison2018}. There have been a few studies exploiting the radio, X-ray, and EUV imaging as well as in-situ observations for understanding the thermodynamic properties of the CMEs.

EUV spectral observations from the ultraviolet coronagraph spectrometers (UVCS), coronal Diagnostic Spectrometer (CDS), and solar ultraviolet measurements of emitted radiation (SUMER) instruments on \textit{SOlar and Heliospheric Observatory} (SOHO) \citep{Domingo1995} have helped us to infer the density, temperature, ionization state and Doppler velocity of CMEs \citep{Raymond2002,Kohl2006}. They have also suggested that CMEs be a loop like structure with helical magnetic field, and probably have higher temperature than the inner corona \citep{Antonucci1997,Ciaravella2000}. The heating rates of CMEs inferred from UVCS observations show that heating of the CME material continues out to 3.5 \textit{R}$_\sun$ and is comparable to the kinetic and gravitational potential energies gained by the CMEs \citep{Akmal2001,Ciaravella2003}. Recently, \citet{Bemporad2010} estimated the physical parameters of plasma (temperature and magnetic field) in pre- and post-shock region using white-light, EUV and radio observations of a fast CME. The combination of density, temperature and ionization state of CMEs constrains their thermal history and have been used to understand the physical processes within CME plasma.

Some internal properties of CMEs is also measured at near and beyond 1 AU from their in-situ observations by instruments onboard Voyagers, Ulysses, Helios, Wind, ACE, and STEREO spacecraft. For example, CMEs in in-situ observations show structure of a spiral magnetic field, low plasma beta, and low temperature than that in the ambient solar wind \citep{Burlaga1981,Richardson1993}. The reason for lower temperature is expansion of CMEs during their propagation as their leading edges usually move faster than the trailing edges and/or pressures inside CMEs are higher than that in the ambient solar wind. Measuring the properties of an individual ICME through the heliosphere is not always possible due to sparse distribution of in situ spacecraft which are rarely radially aligned and further due to difficulty in identification of ICMEs in the solar wind. The ion charge states in ICMEs are often higher implying high temperature (several million K) at their solar source \citep{Lepri2001,Zurbuchen2003}. The thermodynamic evolution of ICMEs have been statistically studied using in-situ observations taken over a range of radial distance from 
0.3 to 30 AU in the heliosphere \citep{Wang2004a,Wang2005a,Liu2005,Liu2006}. These studies have demonstrated the coulomb collision within ICMEs and their moderate expansion compared with theoretical predictions. They also showed that magnetic field decreases faster in ICMEs than in the solar wind, but the density and temperature decreases slower in ICMEs than in the solar wind. This implied that the plasma in the ICMEs must be heated. The radial expansion speed of ICMEs is found to be of the order of the Alfven speed \citep{Jian2008}. Global MHD modeling of ICMEs based on a polytropic approximation to the energy equation has also been done for interpreting the observations \citep{Riley2003,Manchester2004}. However, the polytropic process and turbulence dissipation \citep{Goldstein1995} are intrinsically different and therefore modeling the thermodynamic state of CMEs is difficult.

Although there are few studies addressing the internal states of CMEs, but most of them provide thermodynamics of CMEs only at a certain heliocentric distance and/or at a certain time. Based on a statistical surveys using multi-spacecraft measurements, the polytropic index of CME plasmas was suggested to be around 1.1 to 1.3 
from 0.3 AU to 20 AU which is about constant over solar cycle  \citep{Liu2005,Liu2006} while the polytropic index for solar wind is noted as 1.46 \citep{Totten1995}. Thus, the expansion of an ICME behaves more like an isothermal than adiabatic process which has polytropic index as 1.66. However, the understanding of the continuous evolution of internal state of an individual CME during its heliospheric propagation is still limited.

So far the only attempt to figure out the internal state of an individual CME during its propagation in the outer corona was done by \citet{Wang2009}, (hereafter Paper I), in which a model to reveal the Flux Rope Internal State (FRIS) was developed. Using the model, one could infer the internal forces and thermodynamic properties of CMEs from the coronagraph and heliospheric imagers observations. In their study, they estimated the polytropic index of CME plasma and suggested that
there is continuous heat injection into the CME plasma. They also found that thermal pressure caused the CME expansion while the magnetic force prevented the CME from expansion, and both the forces decrease continuously as CME move away from the Sun. However, there was a mistake in deriving equation 10 in Paper I though it did not affect the final results of their work, and the use of polytropic equation of state (Equation 12 in Paper I) was not proper. Generally, a polytropic process is a thermodynamic process that obeys the relation between fluid's thermal pressure ($p$) and density ($\rho$) by means of an index  ($\Gamma$) described as

\begin{eqnarray}
p=b\rho^\Gamma
\label{polytropic}
\end{eqnarray}
\\
where both $\Gamma$ and $b$ are not constants and change with time. It describes the change of state of any fluid. We emphasize that the value of $b$ was kept constant in Paper I which was incorrect. This is because $b$ should change when a thermodynamic system evolve from one polytropic process to another. 
Further, we use the equation of thermodynamics and derive a few additional parameters, such as absorbed heat, entropy, heating rate and entropy changing rate of the CME. Thus, in the present study, we improve the Flux Rope Internal State (FRIS) model (Section~\ref{fris}), and apply it to the CME of 2008 December 12 to demonstrate what we can learn from the observations with the aid of the model (Section~\ref{appmodel}). A summary and discussion on our study are presented in 
Section~\ref{Resdis}.

\textbf{\section{Flux rope model for CME’s internal state} \label{fris}}

In this model, a CME is treated as an axisymmetric cylinder with the self-similar expansion. The self-similar assumption assumes that the distributions of the quantities along the radial direction in the flux rope frame (with z-axis of a cylindrical coordinates along the flux rope axis) remains unchanged during the propagation of the CMEs. Using this model, one could infer the internal forces and thermodynamic properties of CMEs from the coronagraph observations. The derivation of the improved model is similar to that in Paper I except for some necessary corrections. For the completeness of the paper, we describe the details of the whole derivation below.

\begin{figure}
\begin{center}
\plottwo{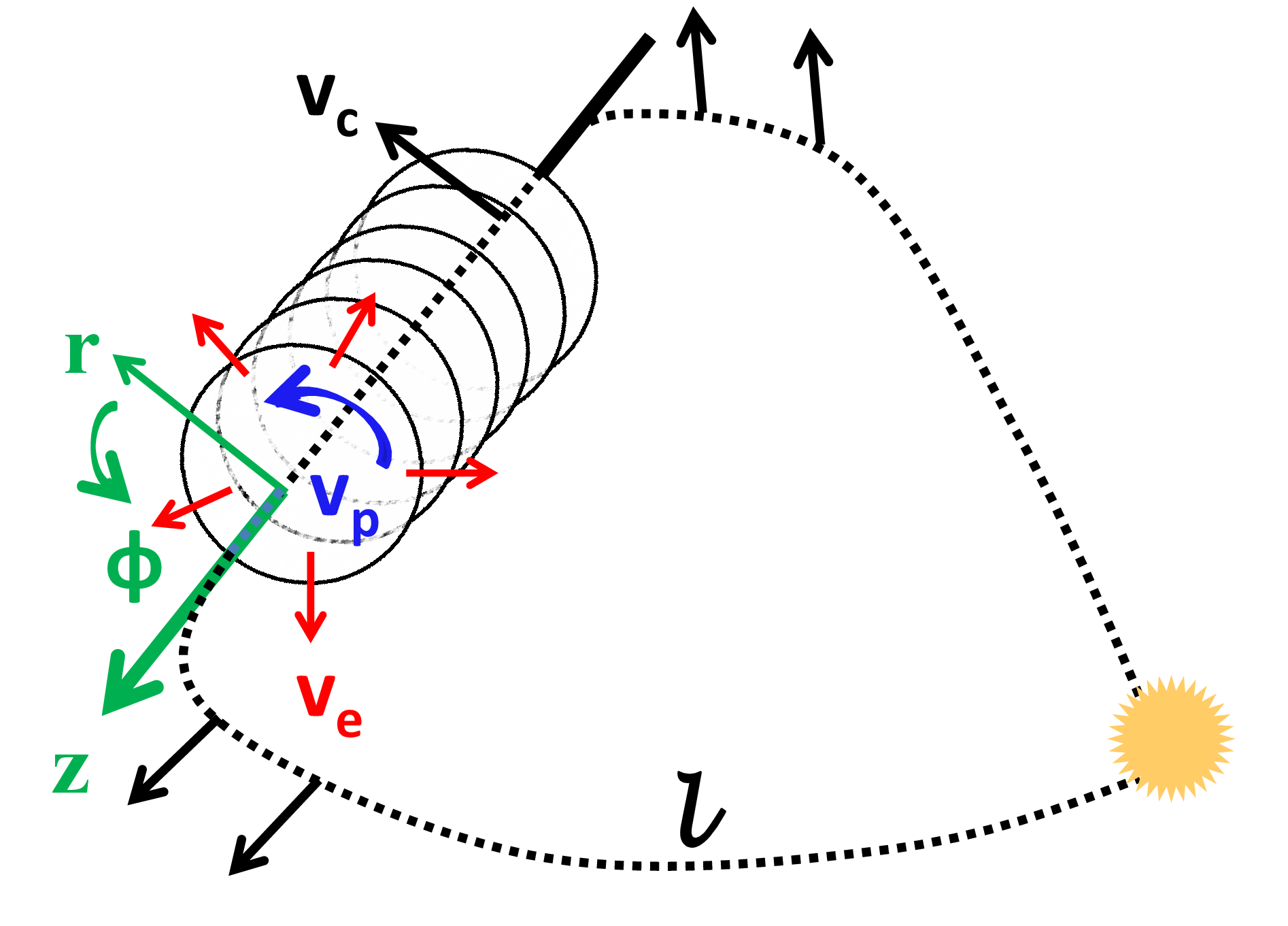}{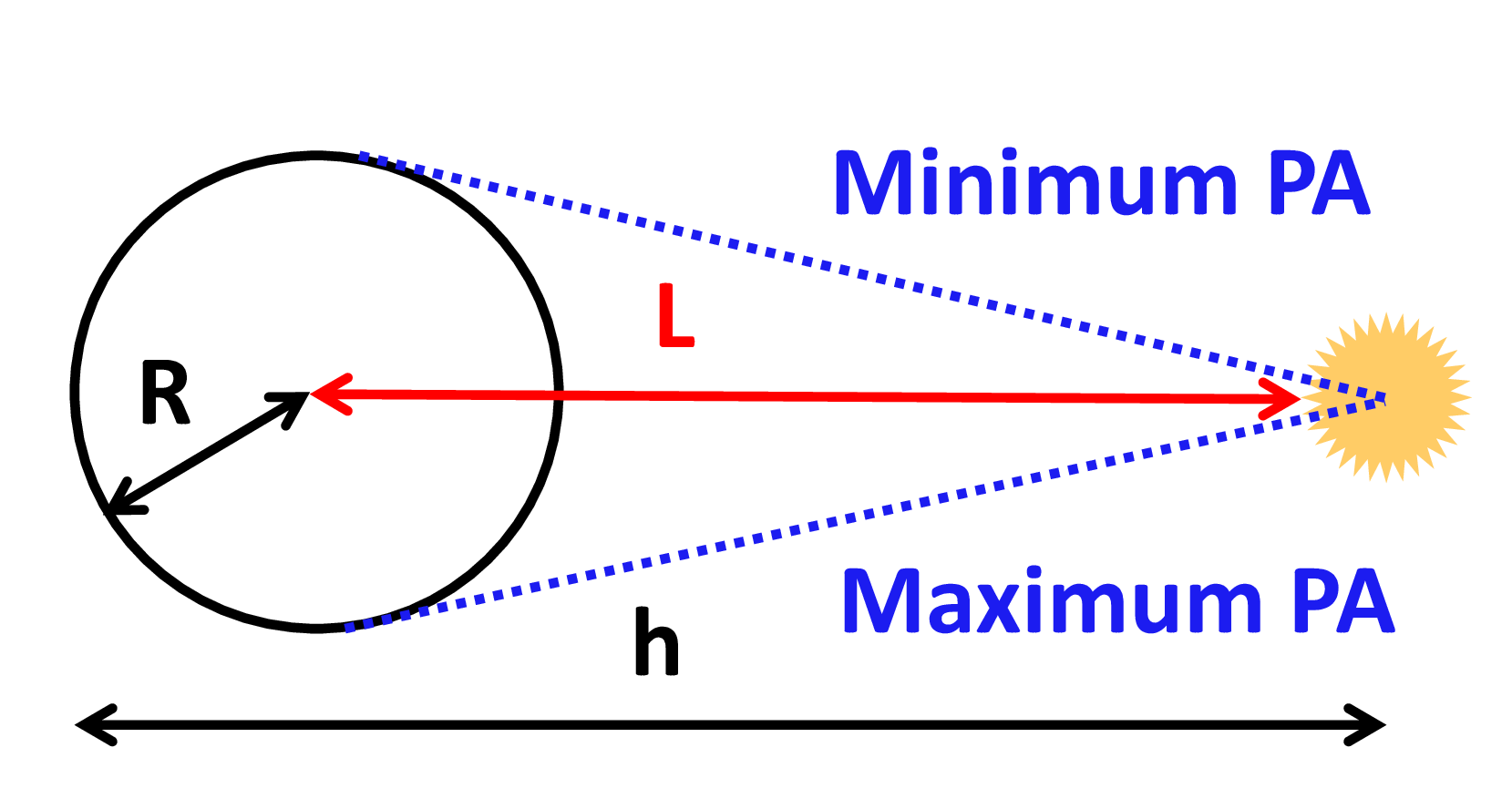}
\caption{Left panel: Schematic picture of a flux-rope CME. The black dashed line indicates the looped axis of the flux rope with the axial length as $l$. A cylindrical
coordinate system (i.e., $r$, $\phi$, $z$) is attached to the the axis of the flux rope (in green). The three types of global motions of flux rope CME is represented. The linear propagation speed ($v_c$), expansion speed ($v_e$), and poloidal speed ($v_p$) are represented with black, red, and blue arrows respectively. The $v_c$ is the speed in a rest reference frame, while $v_e$ and $v_p$ are the speeds at the boundary of the flux rope in a plane perpendicular to the axis of the flux rope. 
Right panel: Schematic picture for the cross-section of the CME flux rope. The radius of the flux rope is $R$, $L$ is the distance from the flux rope axis to the
solar surface, $h$ is the heliocentric distance of the CME leading edge, and the minimum and the maximum position angles are shown with dotted blue lines (adapted from \citealt{Wang2009}). 
\label{fr_cartoon}}
\end{center}
\end{figure}

\textbf{\subsection{Derivation of FRIS model} \label{derivemodel}}

We begin with assumption that the CME to be an axisymmetrical cylindrical flux rope in a local scale, even though in the global scale they may be a loop-like structure with two ends rooted on the surface of the Sun. Therefore, a flux rope can  be treated axisymmetrically in the cylindrical coordinates ($r, \phi, z$) with the origin on the axis of the flux rope (Figure~\ref{fr_cartoon}) and we have $\frac{\partial}{\partial \phi} = \frac {\partial}{\partial z} = 0$. Since the cross-section of the flux rope is a circle, we assume its radius be $R$, which is a time-dependent parameter. Under the consideration of flux rope geometry, the expansion speed of the flux rope's boundary is given by

\begin{eqnarray}
v_e(t) = \frac{dR(t)}{dt}
\end{eqnarray}

and the expansion acceleration at the boundary of the flux rope is

\begin{eqnarray}
a_e(t) = \frac{dv_e(t)}{dt}
\end{eqnarray}

Further, assuming that the flux rope is undergoing a self-similar expansion within the cross section, we can get a dimensionless variable

\begin{eqnarray}
x = \frac{r}{R}
\end{eqnarray}

which is the normalized radial distance from the flux rope axis and independent of time. The $x$=1 is the boundary of the flux rope. This express any self-similar variable as the product of a function of $x$ and a function of $t$. Under this treatment, the radial ($r$) and azimuthal ($\phi$) component of velocity can be expressed as

\begin{eqnarray}
v_r(t,x)  =\frac{d}{dt}(xR)= xv_e(t) \label{velr} \\
v_\phi(t,x)=f_{v_{\phi}}(x)v_p(t)
\end{eqnarray}

where $v_p$ is the poloidal speed at the boundary of the flux rope. The poloidal speed of CME plasma is dependent on both the time and its location from the axis of CME flux-rope. Thus, the radial component of acceleration can be expressed as,

\begin{eqnarray}
a_r(t,x)=\frac{d\mathbf{v}(t,x)}{dt}\cdot\hat{\mathbf{r}}  =x a_e(t) - \frac{v_{\phi}^2(t,x)}{xR}
\label{accrphi}
\end{eqnarray} 

Thus, the first term on the right-hand side of equation~\ref{accrphi} is the acceleration of the radial motion of the plasma, and the second term is the acceleration contributed from the circular/poloidal motion of the plasma.

\textbf{\subsubsection{Conservations of total mass and angular momentum}\label{conservemm}}

Under the assumption of self-similar expansion and considering that magnetic field lines are frozen-in with the plasma flows, the density in the flux rope has a fixed distribution $f_\rho(x)$, and therefore the density ($\rho$) in the flux rope can be expressed as
\begin{eqnarray}
\rho(t,x)=f_\rho(x)\rho_0(t)
\label{dens}
\end{eqnarray} 
 where $\rho_0$ is the average density of the flux rope. The conservation of total mass ($M$) is expressed as

\begin{eqnarray}
M=\int{\rho r dr d\phi dz} = 2\pi l\rho_0 \int_0^R f_\rho(x) r dr = 2 \pi lR^2\rho_0 \int_0^1 f_\rho(x) x dx
\label{mass}
\end{eqnarray} 

where $l$ is the axial length of the flux rope (Figure~\ref{fr_cartoon}). The average density of the flux rope is 

\begin{eqnarray}
\rho_0(t)= \overline{\rho}=\frac{M}{\pi l R^2} 
\label{dens_n1}
\end{eqnarray} 

On combining the equation~\ref{mass} and \ref{dens_n1}, we get $\int_0^1 f_\rho(x) x dx=1/2$. Further, the total angular momentum of the flux rope is expressed as

\begin{eqnarray}
L_A=\int \rho r v_\phi r dr d\phi dz = 2\pi lR^3 \int_0^1 \rho v_\phi x^2 dx = 2\pi l R^3\rho_0(t)v_p(t)\int_0^1 f_\rho(x)f_{v\phi}(x) x^2 dx
\label{momentum}
\end{eqnarray} 

On combining the equation~\ref{dens_n1} and \ref{momentum}, we may derive 

\begin{eqnarray}
v_\phi(t,x) = \frac{k_1 L_A}{M R} f_{v \phi}(x)
\label{velphi}
\end{eqnarray} 
\\
where $k_1 = \frac{1}{2 \int_0^1 f_\rho(x) f_{v\phi} (x) x^2 dx}$.  

\textbf{\subsubsection{Equation of motion}}
We investigate the motion of an arbitrary fluid element using momentum conservation equation (in the frame frozen-in with the moving flux rope) as,

\begin{eqnarray}
\rho \frac{\partial \mathbf{v}}{\partial t} + \rho(\mathbf{v}\cdot \nabla)\mathbf{v} = -\nabla p + \mathbf{j \times B}
\label{eqofmot}
\end{eqnarray} 

where $p$ is the plasma thermal pressure, $\mathbf{B}$ = $(0, B_\phi, B_z)$ is the magnetic induction, and $\mathbf{j}$ = $\frac{1}{\mu_0} (\nabla \times \mathbf{B}$) 
is the current density. Here, the viscous stress tensor ($S$),  gravity force $(\mathbf{F_g})$ , and the equivalent fictitious force ($\mathbf{F_a}$) due to the use of a non-inertial reference frame opposite to the acceleration, are ignored. The the validity of this consideration is discussed in Section~\ref{Resdis}. The 
equation~\ref{eqofmot} can be decomposed into the $\hat{\mathbf{r}}$ and $\hat{\boldsymbol{\phi}}$ components and we have,

\begin{eqnarray}
\hat{\mathbf{r}}&:&  \qquad  \rho \frac{\partial v_r}{\partial t} + \rho(v_r \frac{\partial v_r}{\partial r} - \frac{v_\phi^2}{r}) = - \frac{\partial p}{\partial r} 
+ (\mathbf{j \times B})_r  
\label{eqofmot_r} \\
\hat{\boldsymbol{\phi}}&:&  \qquad \frac{\partial v_\phi}{\partial t} + v_r \frac{\partial v_\phi}{\partial r} + \frac{v_r v_\phi}{r} = 0   
\label{eqofmot_phi}
\end{eqnarray} 

Equation~\ref{eqofmot_phi} means that there is no force acting in azimuthal ($\boldsymbol{\hat{\phi}}$) direction, and the angular momentum is conserved. Substituting the $v_r$ and $v_\phi$ in equation~\ref{eqofmot_r} from equation~\ref{velr} and ~\ref{velphi}, respectively, we obtain 

\begin{eqnarray}
(\mathbf{j \times B})_r = \rho \Big(a_e x - \frac{k_1^2L_A^2 f_{v \phi}^2}{M^2 x R^3} \Big) + \frac{\partial p}{\partial r} 
\label{lorentz}
\end{eqnarray} 

\textbf{\subsubsection{Equation of thermodynamics}}
The laws of thermodynamics (i.e., relationship between heat, work, and the properties of the system in equilibrium) are expressed and fall within the constraints imposed by thermodynamic equations. For ideal gases and reversible processes, the first and second laws of thermodynamics are expressed in the form of 
equation~\ref{tde-1} and ~\ref{tde-2} as 
\begin{eqnarray}
&du=dQ-pd\frac{1}{\rho} \label{tde-1}\\
&ds=\frac{dQ}{T} \label{tde-2}
\end{eqnarray}
\\
where $u$=$\frac{p}{(\gamma-1)\rho}$, and $p=nk(T_p+T_e)=2nkT$. In these formulations, $p$, $u$, $Q$, and $s$ are the thermal pressure, internal energy per unit mass, heat per unit mass and entropy per unit mass, respectively. The $\rho$ is the mass density of protons and $n$ is number density of protons and thus, 
$n$=$\frac{\rho}{m_p}$ where $m_p$ is the proton mass. The $\gamma$ is the adiabatic index which is $\frac{5}{3}$ for monoatomic ideal gases. The $T_p$ and $T_e$ are the temperatures of proton and electron, respectively and $T=\frac{1}{2}(T_p+T_e)$ is the average temperature between the two species. Equation~\ref{tde-2} can be rewritten by using the formulations noted above,  

\begin{eqnarray}
Tds&=&du+pd\frac{1}{\rho} \nonumber\\
&=&\frac{p}{(\gamma-1)\rho}d\ln\left(\frac{p}{\rho^\gamma}\right) \\
\Rightarrow \ln\left(\frac{p}{\rho^\gamma}\right)&=&\frac{(\gamma-1)m_p}{2k}(s+s_0) \nonumber\\
&=&\frac{(\gamma-1)m_p}{2k}s \mathrm{,\ [WLOG\ using\ s\ for\ s+s_0]} \nonumber\\
\Rightarrow p&=&e^{\frac{(\gamma-1)m_p}{2k}s}\rho^\gamma
\end{eqnarray}

Since, $\sigma$=$\frac{(\gamma-1)m_p}{2k}$ is a constant, therefore we can write for our considered flux rope as,
\begin{eqnarray}
p(t,r)&=&e^{\sigma s}\rho^\gamma
\label{polytropic_n1}
\end{eqnarray}

The polytropic process is described in equation~\ref{polytropic} and combinning that with equation~\ref{polytropic_n1}, it is obtained that

\begin{eqnarray}
e^{\sigma s}=b\rho^{\Gamma-\gamma}
\label{polytropic_n2}
\end{eqnarray}

\textbf{\subsubsection{Lorentz force and helical magnetic field structure}} 
We calculate the average Lorentz force ($\overline{f}_{em}$), by integrating equation~\ref{lorentz} over the cross-section of the flux rope and substitute the 
$p$ and $\rho$ from equation~\ref{polytropic_n1} and \ref{dens}, respectively. Thus, we have

\begin{eqnarray}
\overline{f}_{em}&=& \frac{2}{R^2} \int_0^R \mathbf{(j \times B)}_r rdr \nonumber\\
&=&	\frac{k_2 M a_e}{l R^2} - \frac{k_1^2 k_3 L_A^2}{M l R^5} - \frac{k_4 M^\gamma e^{\sigma s}}{l^\gamma R^{\gamma+1}}
\label{lorentz_n1}
\end{eqnarray}																																											
\\
where $k_2$= $\frac{2}{\pi} \int_0^1 f_\rho x^2 dx > 0$, $k_3$= $\frac{2}{\pi} \int_0^1 f_\rho f_{v \phi}^2 dx \ge 0$ and 
$k_4$ = $\frac{2}{\pi^\gamma} \Big[ \int_0^1 f_\rho^\gamma dx - f_\rho^\gamma (1) \Big]$ are integral constants. The first two terms in the right-hand side of equation~\ref{lorentz_n1} are average total force  due to the expansion and poloidal motion, and the third term is the average thermal pressure force

\begin{eqnarray}
\overline{f}_{th} = \frac{k_4 M^\gamma e^{\sigma s}}{l^\gamma R^{\gamma+1}}
\label{thermalf}
\end{eqnarray}

Further, the magnetic field in an axisymmetric cylindrical flux rope can be written as,

\begin{eqnarray}
\mathbf{B} = B_\phi \boldsymbol{\hat{\phi}} + B_z \mathbf{\hat{z}} = \boldsymbol{\nabla} \times \mathbf{A} \nonumber \\
B_\phi = -\frac{\partial A_z}{\partial r} \quad \text{and} \quad B_z = \frac{1}{r} \frac{\partial}{\partial r} \big(r A_\phi \big)
\label{magfphiz}
\end{eqnarray}
  	
Under the self-similar assumption, the magnetic flux is conserved in both $\boldsymbol{\hat{\phi}}$ and $\mathbf{\hat{z}}$ directions, i.e.,

\begin{eqnarray}
\left\{\begin{array}{l}
F_\phi = -l \int_0^R \frac{\partial A_z}{\partial r} dr = l(A_z(0) - A_z(R)) = \text{constant} \nonumber \\ 
F_z = 2\pi \int_0^R \frac{\partial}{\partial r} (r A_\phi) dr = 2 \pi R A_\phi (R) = \text{constant}
\end{array}\right.
\end{eqnarray}

In order to satisfy the self-similar expansion assumption, $A_\phi$ and $A_z$ have to keep their own distributions. Therefore, $A_\phi$ and $A_z$ can be expressed as

\begin{eqnarray}
\left\{\begin{array}{l}
A_\phi = f_{A\phi}(x) A_{\phi0}(t)  \nonumber \\ 
A_z = f_{Az}(x) A_{z0} (t) 
\end{array}\right.
\end{eqnarray}

On combining the above two sets of equations, it is derived that 

\begin{eqnarray}
\left\{\begin{array}{l}
A_\phi(t,x) = \frac{f_{A\phi}(x)}{R}  \\ 
A_z (t,x) = \frac{f_{Az}(x)}{l}
\label{vectorpot}
\end{array}\right.
\end{eqnarray}

Also, it can be proved that the total magnetic helicity ($H_m$) is conserved and therefore we can write as, 

\begin{eqnarray}
H_m = \int \mathbf{B.A} r dr d\phi dz = 2\pi \int_0^1 \Big[ f_{Az} \frac{\partial}{\partial x} (x f_{A\phi}) - x f_{A\phi} \frac{\partial f_{Az}}{\partial x} \Big] dx = \text{constant}
\end{eqnarray}

Combinning equation~\ref{magfphiz} and \ref{vectorpot}, we can calculate the Lorentz force in the flux rope as

\begin{eqnarray}
\mathbf{j  \times B} &=& \frac{1}{\mu_0} (\mathbf{\nabla \times B) \times B} \nonumber \\
&=& -\frac{1}{\mu_0} \left\{ \frac{1}{R^5 x} \frac{\partial}{\partial x} \Big[\frac{1}{x} \frac{\partial}{\partial x} (x f_{A\phi}) \Big]  
                   \frac{\partial} {\partial x} (x f_{A\phi}) + \frac{1}{l^2 R^3 x} \frac{\partial}{\partial x} \Big( x \frac{\partial f_{Az}}{\partial x} \Big)  
									 \frac{\partial f_{Az}}{\partial x} \right\} \mathbf{\hat{r}} \nonumber \\														
\Rightarrow \overline{f}_{em}&=& \frac{2}{R^2} \int_0^R \mathbf{(j \times B)}_r rdr \nonumber \\	
&=& - \frac{k_5}{\mu_0 R^5} - \frac{k_6}{\mu_0 l^2 R^3}
\label{lorentz_n2} 							
\end{eqnarray}
\\		
where $k_5$ = $2 \int_0^1 \frac{\partial}{\partial x} \Big[ \frac{1}{x} \frac{\partial}{\partial x} (x f_{A\phi}) \Big] \frac{\partial}{\partial x} (x f_{A\phi}) dx$ and  $k_6$ = $2 \int_0^1  \frac{\partial}{\partial x} \Big(x \frac{\partial f_{Az}}{\partial x} \Big) \frac{\partial f_{Az}}{\partial x} dx \ge 0$ are both integral constants. The sign of the $k_5$ is determined by the $B_z^2 (R) - B_z^2 (0)$.

\textbf{\subsubsection{Governing equation of motion relating with measurable parameters of the flux rope}} 
From the observations of the present era, it is difficult to precisely measure the axial length ($l$) of a flux rope. However, it can be assumed that axial length of the flux rope has a relationship with distance ($L$) between the axis of the flux rope and the solar surface (Figure~\ref{fr_cartoon}). The $L$ can be measured using the imaging observations of CME/flux rope and therefore we can write, 

\begin{eqnarray}
l = k_7 L
\label{fr_len}
\end{eqnarray}

where $k_7$ is a positive constant. The combination of equation~\ref{lorentz_n1}, \ref{lorentz_n2} and \ref{fr_len} results in the final equation as

\begin{eqnarray}
a_e - c_1 R^{-3} = c_2 L R^{-3} + c_3 L^{-1} R^{-1} + \lambda L^{1-\gamma} R^{1-\gamma}
\label{govern}
\end{eqnarray}
\\
where the coefficients $c_{1-3}$ are constants, and $\lambda$ is a time-dependent variable. They are given as
 
\begin{eqnarray}
\left\{\begin{array}{l}
c_1 = \frac{k_1^2 k_3 L_A^2}{k_2 M^2} \ge 0  \\
c_2 = \frac{- k_5 k_7}{\mu_0 k_2 M}  \\
c_3 = \frac{- k_6}{\mu_0 k_2 k_7 M}  \le 0
\label{const_1to3}
\end{array}\right.
\end{eqnarray}

\begin{eqnarray}
\lambda (t) = c_0 e^{\sigma s(t)}
\label{lambda_td}
\end{eqnarray}

where coefficient $c_0$ is a constant and $c_0$ =$\frac{k_4 M^{\gamma-1}}{k_2 k_7^{\gamma-1}}$. The left-hand side of equation~\ref{govern} describes the motion of the plasma in the flux rope. Its first and second term gives the acceleration due to the radial motion (i.e., expansion or contraction) and the acceleration due to the poloidal motion, respectively. 
In the right-hand side of equation~\ref{govern}, the first two terms represents the contribution from the Lorentz force and the third term represents the contribution from the thermal pressure force. The expansion acceleration ($a_e$), distance between the axis of the flux rope and solar surface ($L$) and radius of the flux rope ($R$) in equation~\ref{govern} can be measured from the observations of flux rope CME. Thus, the unknown coefficients $c_{1-3}$ can be obtained by fitting the equation to the measurements as long as we have an additional constraint for the variable $\lambda$, which will be continued in Section~\ref{addconst}.

\textbf{\subsection{Physical parameters of the internal state of CME} \label{internparam}}

\textbf{\subsubsection{Lorentz, thermal and centrifugal force}}

The average Lorentz force ($\overline{f}_{em}$) and thermal pressure force ($\overline{f}_{th}$) can be expressed by substituting the value of $l$ from 
equation~\ref{fr_len} in the equation~\ref{lorentz_n2} and \ref{thermalf}, respectively. Thus,

\begin{eqnarray}
\overline{f}_{em} &=& \frac{k_2 M}{k_7} \big( c_2 R^{-5} + c_3 L^{-2} R^{-3} \big) \label{emforce} \\
\overline{f}_{th} &=& \frac{k_2 M}{k_7} \big( \lambda L^{-\gamma} R^{-\gamma-1} \big) \label{thforce}
\end{eqnarray}

The ratio between the average Lorentz force ($\overline{f}_{em}$) and thermal pressure force ($\overline{f}_{th}$) can be expressed as
\begin{eqnarray}
\frac{\overline{f}_{em}}{\overline{f}_{th}} = \frac{c_2 L^\gamma R^{\gamma-4} + c_3 L^{\gamma-2} R^{\gamma-2}}{\lambda} \nonumber
\end{eqnarray}

Further, the poloidal motion should be maintained by an inward centripetal force, which is a part of the sum of the Lorentz force and thermal pressure gradient force. Therefore, the expression of average outward centrifugal force ($\overline{f}_{p}$) due to the poloidal motion, can be derived using the second term in left-hand side of equation~\ref{govern} as,

\begin{eqnarray}
\overline{f}_{p} &=&  \frac{k_2 M}{k_7} \big(c_1 R^{-5} L^{-1} \Big)  \label{poforce} 
\end{eqnarray}

\textbf{\subsubsection{Density, temperature and thermal pressure}}
On combining the equation~\ref{dens_n1} and ~\ref{fr_len}, the average proton mass density ($\overline{\rho}$) and proton number density ($\overline{n}_p$) can be expressed as 

\begin{eqnarray}
\overline{\rho}  &=& \frac{M}{k_7 \pi} \big(L R^2 \big)^{-1} \label{dens_n2} \\
\overline{n}_p  &=& \frac{M}{k_7 \pi m_p} \big(L R^2 \big)^{-1}
\label{dens_nn2}
\end{eqnarray}

The average thermal pressure is given as

\begin{eqnarray}
\overline{p}(t) &=& \frac{2}{R^2} \int_0^R p r dr  \nonumber \\
&=& \frac{k_2 k_8 M}{k_4 k_7} \lambda (L R^2)^{-\gamma}
\label{thpres}
\end{eqnarray}
\\
where $k_8$ = $\frac{2}{\pi^\gamma} \int_0^1 f_\rho^\gamma x dx > 0$ is a constant and the average temperature ($\overline{T}(t)$) is given as

\begin{eqnarray}
\overline{T}(t) &=& \frac{m_p}{2k} \frac{\overline{p}}{\overline{\rho}}  \nonumber \\
&=&\frac{\pi \sigma k_2 k_8}{(\gamma-1)k_4} \lambda \big ( L R^2 \big)^{1-\gamma} 
\label{temp}
\end{eqnarray}.

\textbf{\subsubsection{Polytropic index}}
The polytropic process can describe multiple expansion and compression processes which include heat transfer. The evolution of the polytropic index can be obtained from the combination of equation~\ref{polytropic_n2} and ~\ref{lambda_td} as follows

\begin{eqnarray}
\lambda &=& c_0 b \rho^{\Gamma-\gamma}  \nonumber \\
\Gamma &=& \gamma + \frac{\ln \lambda - \ln \big(c_0 b \big) }{\ln \rho} 
\label{polytropic_n3}
\end{eqnarray}

Assume the thermodynamic process is quasi-static between each two measurement points within a relatively short period, i.e., between the instants $t$ and 
$t + \Delta t$. This implies that the values of $\Gamma$ and $b$ in equation~\ref{polytropic_n3} are nearly constant during that time interval.  Then the 
equation~\ref{polytropic_n3} can be solved as

\begin{eqnarray}
&&\left\{\begin{array}{l}
\Gamma (t) = \gamma + \frac{\ln \lambda (t) - \ln \big(c_0 b (t) \big) }{\ln \rho (t)}   \nonumber \\  
\Gamma (t) \approx \gamma + \frac{\ln \lambda (t + \Delta t) - \ln \big(c_0 b (t) \big) }{\ln \rho (t + \Delta t)}  \nonumber \\  
\end{array}\right. \\ \nonumber
&&\Rightarrow \Gamma (t) \approx \gamma + \frac {\ln \frac{\lambda(t)}{\lambda(t+\Delta t)}} {\ln \frac{\rho (t)}{\rho (t+ \Delta t)}}  \nonumber \\
&& = \gamma+\frac {\ln \frac{\lambda(t)}{\lambda(t+\Delta t)}} {\ln \left\{\frac{L (t+\Delta t)}{L (t)} \Big[ \frac{R(t+\Delta t)}{R (t)} \Big]^2 \right\}}
\label{polytropic_f}
\end{eqnarray}

\textbf{\subsubsection{Rate of change of entropy and heating rate}}
Further, we derive the expression for the rate of change of entropy ($\frac{ds}{dt}$) per unit mass from the expression of $\lambda$ as in equation~\ref{lambda_td},  
 
\begin{eqnarray}
\frac{d\lambda}{dt}  =\sigma \lambda \frac{ds}{dt}  \nonumber \\
\frac{ds}{dt} = \frac{1}{\sigma \lambda} \frac{d\lambda}{dt}
\label{rtofentropy}
\end{eqnarray}

Now using the second law of thermodynamics as in equation~\ref{tde-2}, we derive the expression for the average heating rate ($\overline{\kappa}$) per unit mass. This is given as,

\begin{eqnarray}
\overline{\kappa}(t) &=& \frac{d Q(t)}{dt} = \overline{T} \frac{ds}{dt} \nonumber \\
&=& \frac{\pi k_2 k_8}{(\gamma-1) k_4} \big( LR^2 \big)^{1-\gamma} \frac{d\lambda}{dt}
\label{heatingrate}
\end{eqnarray}                                                                                                                                      

\textbf{\subsubsection{Total energies of the CME plasma}}

The total thermal energy ($E_i$) and total magnetic energy ($E_m$) of plasma associated with the flux rope of CME can be given as

\begin{eqnarray}
E_i &=& 2 \pi l \int_0^R \frac{p}{\gamma-1} r dr  \nonumber \\
&=& \frac{\pi k_2 k_8 M}{(\gamma-1) k_4} \lambda (L R^2)^{1-\gamma}    \label{energy_th}  \\
E_m &=& 2 \pi l \int_0^R \frac{B^2}{2 \mu_0} r dr \nonumber \\
&=&  \frac{\pi k_9}{\mu_0 k_7} L^{-1} + \frac{\pi k_7 k_{10}}{\mu_0} L R^{-2}    \label{energy_em}
\end{eqnarray} 
\\
where $k_9$ = $\int_0^1 x \big(\frac{\partial f_{Az}}{\partial x} \big)^2 dx \ge 0$ and  $k_{10}$ = $\int_0^1 \frac{1}{x} \big[ \frac{\partial}{\partial x} (x f_{A\phi}) \big]^2 dx \ge 0$ are both integral constants.

\textbf{\subsection{Additional constraint for derivation of the unknown coefficients and the variable} \label{addconst}}

The governing equation of motion for the expanding flux rope of the CME as given in equation~\ref{govern} can be expressed as

\begin{eqnarray}
\lambda = L^{\gamma-1} R^{\gamma-1} (a_e - c_1 R^{-3} -c_2 L R^{-3} -c_3 L^{-1} R^{-1}) 
\label{govern_n1}
\end{eqnarray}

From equation~\ref{govern_n1}, we can write as 

\begin{eqnarray}
\frac{d\lambda}{dt} &=& L^{\gamma-1} R^{\gamma-1} \Big[ \big\{ \frac{da_e}{dt} + (\gamma-1) (a_e v_e R^{-1} + a_e v_c L^{-1}) \big\}  \nonumber \\
&  &                + c_1 \big\{ (4-\gamma) v_e R^{-4} + (1-\gamma) v_c L^{-1} R^{-3}  \big\}   \nonumber \\
&  &                +c_2 \big\{ (4-\gamma) v_e L R^{-4} - \gamma v_c R^{-3} \big\}  \nonumber \\
&  &                +c_3 \big\{ (2-\gamma) (v_e L^{-1} R^{-2}  + v_c L^{-2}  R^{-1} ) \big\} \Big] 
\label{dgovern_n1}
\end{eqnarray}

In equation~\ref{govern_n1}, $R$, $L$ and $a_e$ are all the known parameters that can be obtained from the measurements using coronagraphic observations of the CMEs, and the others are unknown parameters. However, only equation~\ref{govern_n1} is not sufficient to obtain the unknown coefficients $c_{1-3}$ and the variable 
$\lambda$. Therefore, we need an additional constraint on $\lambda$ to determine the unknown coefficients. For this purpose, we assume that irrespective of the any heating mechanism in the CMEs,  the heating rate ($\overline{\kappa}$) per unit mass of the CME plasma may be equivalently expressed as the result of heat flow. Thus,

\begin{eqnarray}
\overline{\kappa} = k_{11} \frac{T_a - \overline{T}}{\overline{\rho} L^2}
\label{heatflow}
\end{eqnarray}
\\
where $k_{11}$ is an unknown positive constant and $T_a$ could be treated as equivalent temperature of the ambient solar wind around the CME base, which serves as a heat source for the CME plasma. The equation~\ref{heatflow} is analogous to the expression of heat flow per unit mass, $\frac{1}{\rho}\nabla\cdot\ \mathbf{q}$, in which $\mathbf{q}$ is the heat flux for conduction, and proportional to $\nabla T$.  Substituting 
$\overline{\rho}$, $\overline{T}$ and $\overline{\kappa}$ in equation~\ref{heatflow} from the equation~\ref{dens_n2}, ~\ref{temp} and ~\ref{heatingrate}, we get 

\begin{eqnarray}
(L R^2)^{\gamma-1} = c_4 L R^{-1} \frac{d\lambda}{dt} + c_5 \lambda
\label{govern_n2}
\end{eqnarray}
\\
where the coefficients $c_4$ and $c_5$ are constants which are given as

\begin{eqnarray}
\left\{\begin{array}{l}
c_4 = \frac{k_2 k_8 M}{(\gamma-1) k_4 k_7 k_{11} T_a}  \\
c_5 = \frac{\pi \sigma k_2 k_8}{(\gamma-1) k_4 T_a}  \\
\end{array}\right.
\end{eqnarray}

From the value of $c_4$, we can write the expression for the temperature ($T_a$) of the ambient solar wind around the CME base as

\begin{eqnarray}
T_a  = \frac{\pi\sigma k_2 k_8}{(\gamma-1) k_4 c_5}  
\end{eqnarray}

On substituting the value of $\lambda$ and $\frac{d\lambda}{dt}$ from equation~\ref{govern_n1} and ~\ref{dgovern_n1} in equation~\ref{govern_n2}, we can finally write 

\begin{eqnarray}
(L R^2)^{\gamma-1} &=& L^{\gamma-1} R^{\gamma-1} \Big[ c_5 a_e + \big\{(\gamma-1) c_4 a_e v_c - c_3 c_5 L^{-1} + c_4 \frac{da_e}{dt} L \big\} R^{-1} \nonumber \\
& & + \big\{ (2-\gamma) c_3 c_4 v_c L^{-1} + (\gamma-1) c_4 a_e v_e L \big\} R^{-2} \nonumber \\
& & + \big\{ (2- \gamma) c_3 c_4 v_e - c_2 c_5 L - c_1 c_5 \big\} R^{-3} \nonumber \\
& & + \big\{ (1-\gamma) c_1 c_4 v_c - \gamma c_2 c_4 v_c L \big\} R^{-4} \nonumber \\
& & + \big\{ (4-\gamma) c_1 c_4 v_e L + (4-\gamma) c_2 c_4 v_e L^2 \big\} R^{-5} \Big] \nonumber \\
\label{govern_n1n2}
\end{eqnarray}
\\
where $v_c$, $v_e$, $a_e$ and $\frac{da_e}{dt}$ are the propagation speed, expansion speed, expansion acceleration and rate of change of expansion acceleration of the flux rope, respectively. Now the unknown constants $c_{1-5}$ serve as fitting parameters that can be obtained by fitting the equation~\ref{govern_n1n2} to the measurements of distance between the axis of the flux rope and the solar surface ($L$) and radius of the flux rope ($R$) and their derivatives. Once the constants are known, then the variable $\lambda$ can be calculated from equation~\ref{govern_n1}.  Further, the goodness-of-fit can be evaluated by the relative error in the quantity kept on the left-hand side of equation~\ref{govern_n1n2}, i.e., $(LR^2)^{\gamma-1}$.

If the flux rope of the CME is evolving in such as way that its aspect ratio ($\frac{R}{L}$) is constant then we can express $R = k_{12} L$, where $k_{12}$ is a positive constant. It implies that the angular width of the flux rope or CME will not change during the its heliospheric propagation. Consequently, the distance of the leading edge of the flux rope ($h$) is therefore given by, $h = (1+\frac{1}{k_{12}}) R = (1 + k_{12}) L$, and speed and acceleration of the leading edge are 
$v_{h} = (1 + \frac{1}{k_{12}}) v_{e} = (1 + k_{12}) v_c$ and $a_h = (1+\frac{1}{k_{12}}) a_{e} = (1 + k_{12}) a_c$, respectively. Therefore, for the flux rope having constant aspect ratio, the equation~\ref{govern_n1n2} can be expresed as equation~\ref{govern_n1n2fcar} where $h$ is used instead of $R$ or $L$, because it is easier to measure $h$ from the imaging observations of the CMEs.

\begin{eqnarray}
\Big( \frac{k_{12}}{1 + k_{12}} \Big)^{\gamma+1} h^\gamma &=&  \Big[ \Big(\frac{k_{12}}{1 + k_{12}} \Big)^3 \Big( c_5 a_h + \frac{c_4}{k_{12}}  
                                         \frac{da_h}{dt}\Big) \Big] h  - \Big[ c_5 \Big(\frac{c_2}{k_{12}} + c_3 k_{12} \Big)\Big] h^{-1}    \nonumber \\
& & - \Big[ c_1 c_5 \Big(\frac{1+k_{12}}{k_{12}} \Big) + 2 (\gamma-2) \frac{c_4}{k_{12}} \Big (\frac{c_2}{k_{12}} +c_3 k_{12} \Big) v_h \Big] h^{-2}  \nonumber \\
& & + \Big[(5- 2 \gamma) c_1 \frac{c_4}{k_{12}} \Big(\frac{1+k_{12}}{k_{12}} \Big) v_h \Big] h^{-3}
\label{govern_n1n2fcar}
\end{eqnarray}

Further, we know that for a force-free flux rope, $\mathbf{j \times B} = 0$. Therefore, equation~\ref{lorentz_n2} and \ref{const_1to3} imply that the constants 
$k_5$ = $k_6$ = 0 and $c_2$ = $c_3$ = 0. Thus, the equation~\ref{govern_n1n2} and ~\ref{govern_n1n2fcar} can further be simplified under a constraint of force-free flux rope of the CME. However, for the present study, we use the general form of the formulations as expressed in equation~\ref{govern_n1n2} to fit the kinematic data which is explained in Section~\ref{intparam}. Up to now, we have obtained all the internal dynamic and thermodynamic parameters that could be derived by our FRIS model. The summary of the derived physical parameters and the constants are given in Table~\ref{summ}.

\textbf{\section{Application of the model to the CME of 2008 December 12} \label{appmodel}}

\begin{figure}
\begin{center}
\includegraphics[angle=0,scale=0.50]{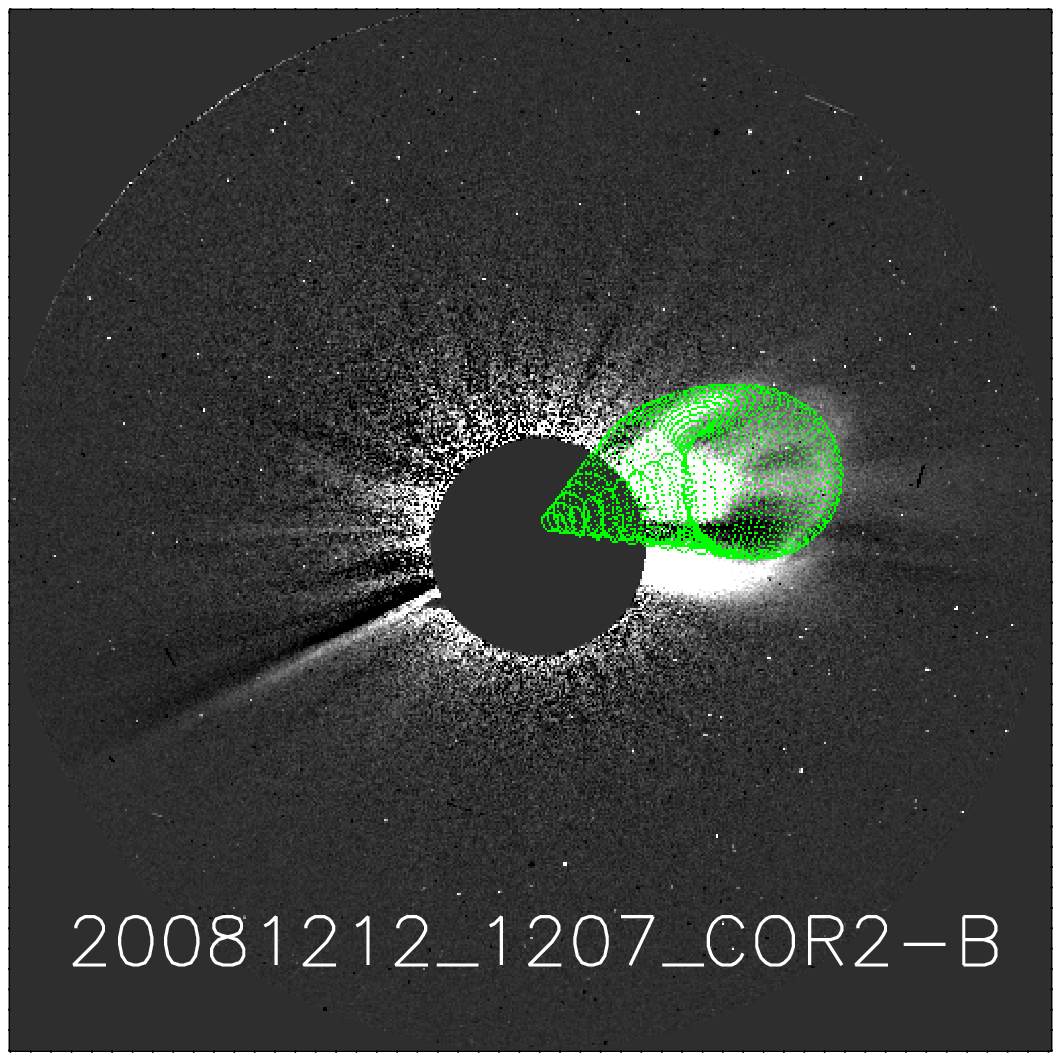}
\includegraphics[angle=0,scale=0.50]{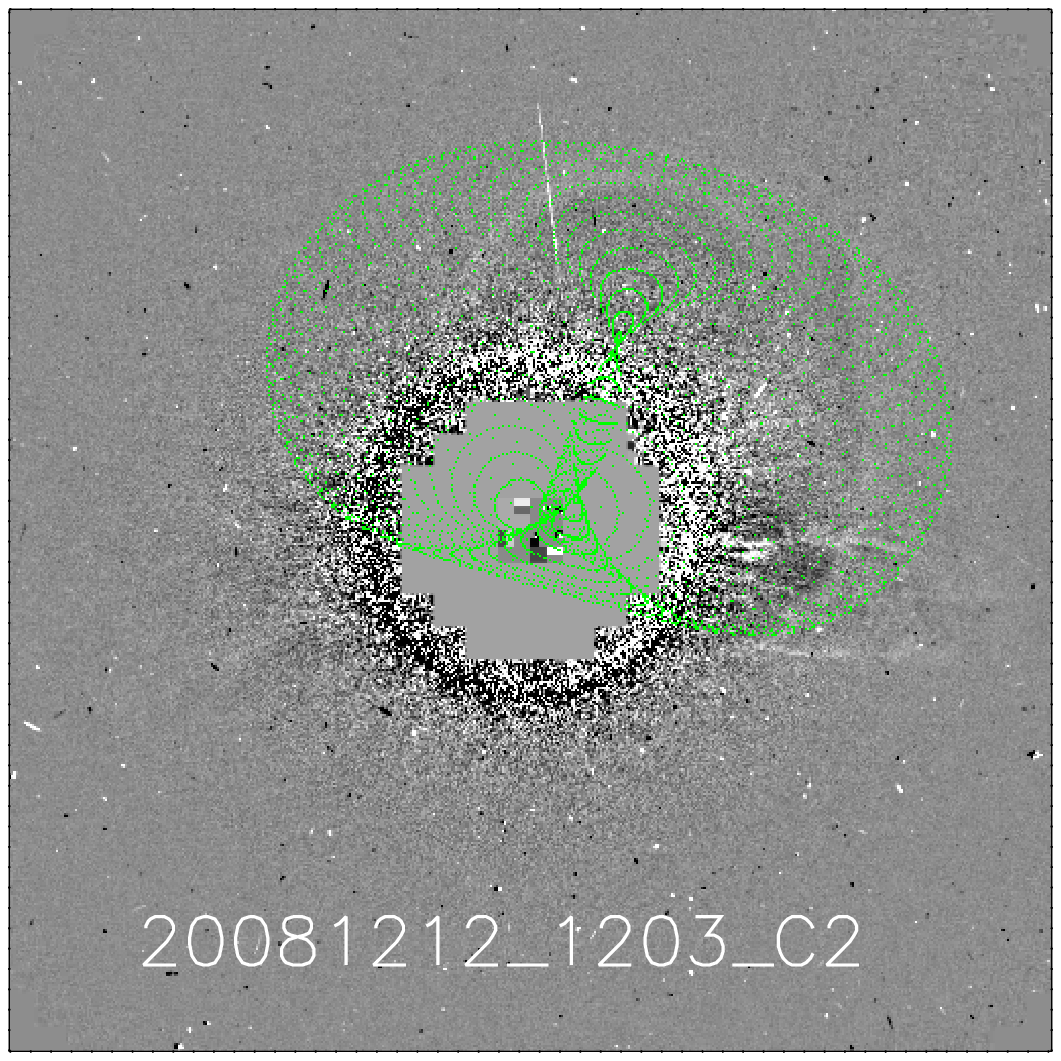}
\includegraphics[angle=0,scale=0.50]{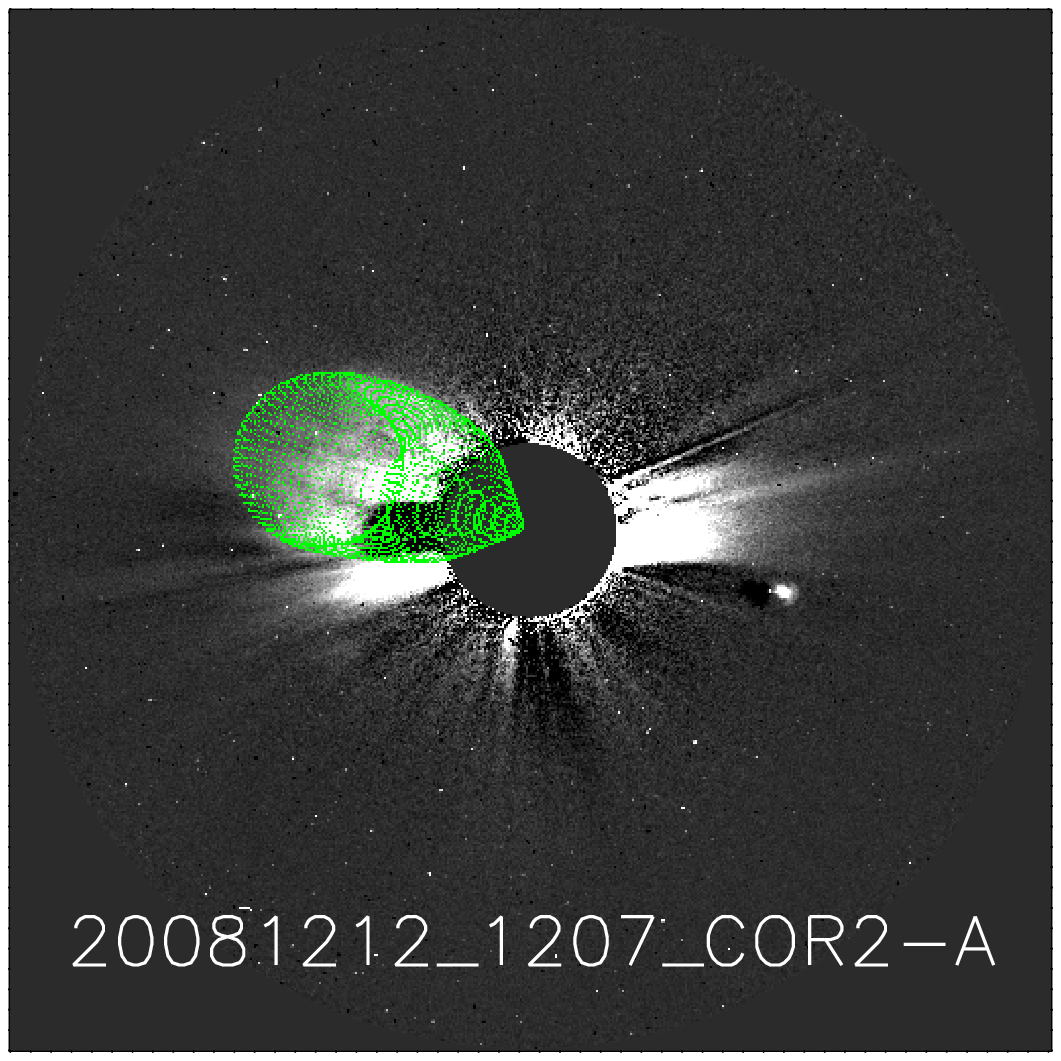}
\caption{The GCS model wire-frame with green is overlaid on the images of 2008 December 12 CME. The triplet of concurrent images are taken from \textit{STEREO}/COR2-B (left), SOHO/LASCO-C2 (middle), and \textit{STEREO}/COR2-A (right) around 12:07 UT on 2008 December 12.}
\label{GCS}
\end{center}
\end{figure}

\textbf{\subsection{Observations and measurements from coronagraphs} \label{obs}}

The CME is well observed by the SOHO/LASCO, Sun-Earth Connection Coronal and Heliospheric Investigation (SECCHI; \citealt{Howard2008}) coronagraphs and the Heliospheric Imagers (HIs) on board STEREO. This CME was observed in SECCHI/COR1-A images at 04:35 UT in the NE quadrant and in SECCHI/COR1-B at 04:55 UT in the NW quadrant. SOHO/LASCO observed this CME as a partial halo with an angular width of 184$\arcdeg$ and a linear speed of 203 km s$^{-1}$. This CME was tracked in LASCO-C3 images out to 12 \textit{R}$_\odot$ where its quadratic speed was measured as 322 km s$^{-1}$. The CME was associated with a prominence eruption that started at 03:00 UT in the NE quadrant observed in SECCHI/EUVI-A 304 {\AA} images. Prominences are always above polarity inversion lines and is believed to be supported by surrounding coronal magnetic field, which may be developed into a flux rope and therefore a CME when an eruption takes place \citep{Demoulin1989,Martens1990,Gibson2006,Okamoto2008,Vemareddy2017}. Thus, the orientation of a prominence usually indicates the axis of the CME flux rope. According to this picture, the CME was approximately viewed from axis in STEREO-A, and viewed aside in STEREO-B. We tracked the CME from the inner corona to outer corona and noted that it took about 
11 hrs for leading edge to travel from 1.4 to 15 \textit{R}$_\odot$ which implies its slow speed.

As it has been described in section~\ref{derivemodel}, we treat a CME as a flux rope, and the input parameters of our model are the radius ($R$) of cross-section the CME’s, the distance ($L$) between the axis of the flux rope and the solar surface and their derivatives. There are various approaches to get $R$ and $L$ for a CME. In earlier studies of \citet{Wang2009}, the $R$ and $L$ for the 2009 October 8 CME is derived by measuring the heliocentric distance ($h$) of the leading edge of that CME as well as its span angle ($\Delta\theta$). Such an approach is suitable for CMEs having unambiguous boundaries and viewed along the axis. The front of the 2008 December 12 CME studied here is diffusive (or fainter) and it is difficult to precisely measure its leading edge and span angle, although it did have an axial-view in STEREO-A. Therefore, we used the Graduated Cylindrical Shell (GCS) model \citep{Thernisien2009,Thernisien2011} to recover the three-dimensional geometry (including $R$ and $L$) of the CME from the projected images of the CME. The GCS forward modeling method have been frequently used to determine the 3D kinematics of the flux rope CMEs \citep{Lynch2010,Gui2011,Bosman2012,Vourlidas2013,Wang2014,Mishra2015a,Mishra2016,Wang2018}.

The model has nine free parameters (refer to Table 1 of \citealt{Thernisien2006}), in which six parameters determine the CME shape projected on the plane-of-sky. These parameters called as geometric parameters, are the longitude ($\phi$), latitude ($\theta$), height (or heliocentric distance) of leading edge ($h$), aspect ratio ($a$), tilt angle and half angle between the two legs of the flux rope. As per our experience of using GCS model, a set of reasonable initial values of the free
parameters are crucial for the model to get the correct values of the parameters. Except that we manually adjust the initial values to make the modeled flux rope match the CME’s 2D projected geometry as much as possible. We also used observations of the source region of the CME to constrain the tilt angle of the flux rope, as the tilt angle should be roughly aligned with the polarity inversion line associated with the CME.

\begin{figure}
\begin{center}
\includegraphics[angle=0,scale=0.75]{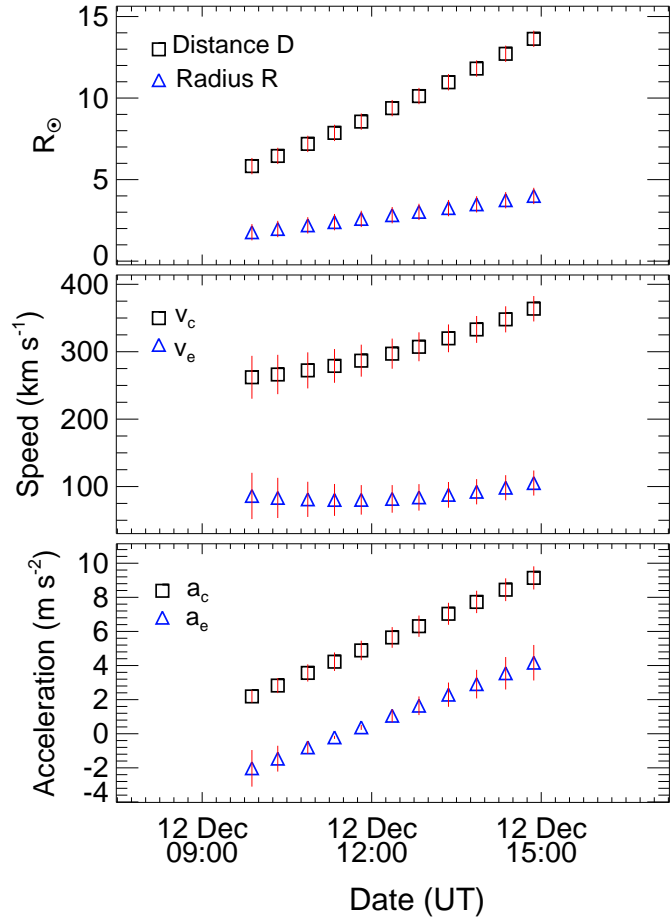}
\caption{Top panel: The measurements of the heliocentric distance ($D$) of the center of flux-rope associated CME and its radius ($R$). Middle panel: The propagation speed ($v_c$) and expansion speed ($v_e$) derived from by taking the derivative of $D$ and $R$, respectively. Bottom panel: The propagation acceleration ($a_c$) and expansion acceleration ($a_e$) derived from by taking the derivative of $v_c$ and $v_e$, respectively. The vertical lines at each data points shows the error bars derived by arbitrarily assuming an error of 0.5 \textit{R}$_\odot$ in the measurements of $D$ and $R$ from coronagraphs observations.}
\label{kinem}
\end{center}
\end{figure}

We applied the GCS forward fitting model to the contemporaneous images of the CMEs obtained from the SECCHI/COR2-B, SOHO/LASCO-C3, and SECCHI/COR2-A coronagraphs as shown in the images overlaid with the fitted GCS wireframed contour (Figure~\ref{GCS}). The obtained longitude, latitude, tilt angle, aspect ratio and half angle 
are 6.7$\arcdeg$, 10.6$\arcdeg$, -15$\arcdeg$ (i.e., clockwise from the ecliptic plane), 0.29 and 15$\arcdeg$, respectively at the height of 11.4 R$_\odot$. The 2008 December 12 CME has been studied earlier by several researchers using STEREO COR and HI observations as well as MHD simulation \citep{Davis2009,Byrne2010,Liu2010a,Poomvises2010,Gui2011,Mishra2013}. Based on the tracking of different features of this CME in HI observations, \citet{Davis2009} estimated the propagation direction of the CME to be at maximum -15$\arcdeg$ to 15$\arcdeg$ away from the Sun-Earth line. In another study, using the GCS model, \citet{Liu2010a} showed that CME has a propagation direction of about 10$\arcdeg$ west of the Sun-Earth line and 8$\arcdeg$ above the ecliptic plane, with a tilt angle of about -53$\arcdeg$ clockwise from the ecliptic plane. The CME is found to show deflection as it propagated away from the Sun and a physical explanation of this has been addressed in \citet{Gui2011}. The GCS fitted parameters in \citet{Poomvises2010} are within 10\% uncertainties of those estimated in the present study, except for the tilt angle and aspect ratio of the CME which are found to be 87$\arcdeg$ and 0.10, respectively, in their study. Using tie-pointing method of 3D reconstruction \citep{Inhester2006,Thompson2009} on the COR data, \citet{Mishra2013} estimated the longitude and latitude of the CME as 5$\arcdeg$ at 15 R$_\odot$. Thus, the values of fitted parameters for the CMEs, except tilt angle which usually has large uncertainty, in the present study are in fair agreement to those estimated in earlier studies. For the purpose of using the FRIS model, the radius ($R$) and the distance ($L$) is derived from the precise measurement of the height ($h$) and aspect ratio ($a$) of the flux rope. The radius of the flux rope is given by $R$ = $\frac{a}{1+a} h$, and its distance as  $L$ = $h$ - $R$ - 1 $\textit{R}_\odot$. We also defined the heliocentric distance of the CME center as $D$ = $L$ + 1 \textit{R}$_\odot$.

We tracked the CME from $D$ = 2 to 14 \textit{R}$_\odot$, and during this interval its radius expanded to about 4 \textit{R}$_\odot$, the propagation speed was slowly accelerated to about 350 km s$^{-1}$, the expansion speed reached to about 100 km s$^{-1}$, and both the propagation and expansion accelerations are positive with peaks at about $D = 2.5 $\textit{R}$_\odot$. The CME shows an impulsive expansion acceleration and then a deceleration phase in the COR1 FOV which continues up to 
COR2 FOV and then again a slight increase in the accelerations is observed. Since, our FRIS model assumes that CME/flux rope is undergoing a self-similar expansion, it is not surprising that the model would not suit for deriving the internal parameters of the CME near the Sun in COR1 FOV. This issue would further be discussed in Section~\ref{intparam} and \ref{Resdis}. Therefore, we have applied the FRIS model on the measurements of CME derived from the COR2 observations alone. Figure~\ref{kinem} presents the time variation of heliocentric distance ($D$), radius ($R$), propagation speed ($v_c$) and expansion speed ($v_e$), propagation acceleration ($a_c$) and expansion acceleration ($a_e$) of the CME of 2008 December 12 observed in COR2 FOV.

\textbf{\subsection{Internal thermodynamic parameters of the CME} \label{intparam}}

The obtained kinematic parameters shown in Figure~\ref{kinem} are used as inputs in the FRIS model. We fit equation~\ref{govern_n1n2} and determine the value of unknowns in the equation. Thus, we fit the general equation, and do not use its simplistic form for force-free flux-rope, which governs the motion of CME as conceptualized in the FRIS model. For this purpose, firstly, the parameter $(LR^2)^{\gamma-1}$ kept in left-hand side of the equation~\ref{govern_n1n2} is determined directly from the measured data of the CME. Then we fit the measured results of $(LR^2)^{\gamma-1}$ with model derived expression as in right-hand side of equation~\ref{govern_n1n2} to find the best set of unknown model fitting parameters (i.e., $c_{1-5}$) which match the measured results in a least-squares sense. The fitting was performed using $MPFITFUN$ routine of IDL which can fit a user-supplied model to a set of user-supplied measured data \citep{Markwardt2009}. Secondly, we input the obtained values of $c_{1-5}$ and observed kinematic parameters in equation~\ref{govern_n1} and ~\ref{dgovern_n1} and determine the value of $\lambda$ and $\frac{d\lambda}{dt}$, respectively. Finally, once the values of $\lambda$, its derivative and $c_{1-5}$ are known, we estimate several internal thermodynamic parameters of the CME.

\begin{figure}
\begin{center}
\includegraphics[angle=0,scale=0.60]{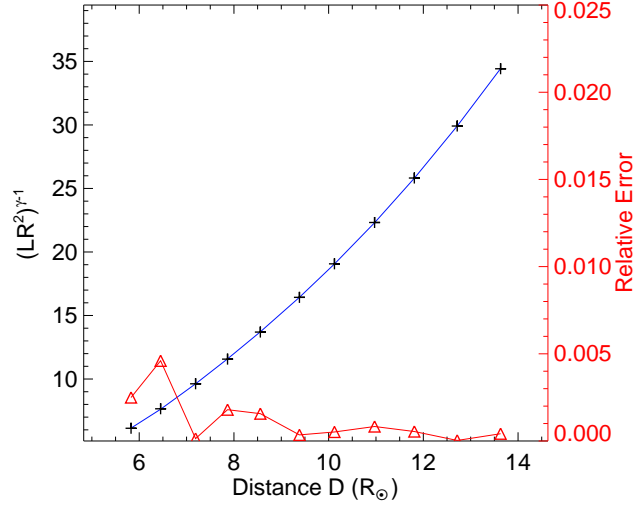}
\caption{The profile of $(LR^2)^{\gamma-1}$ from the measurements (black), the modeled result for this parameter (blue), and the relative error (red) is shown.}
\label{errfit}
\end{center}
\end{figure}

Doing fitting as explained above, the goodness-of-fit is shown in Figure~\ref{errfit}. In the figure, the plus symbols present the quantity  $(LR^2)^{\gamma-1}$, defined as $Q_m$, directly derived from the measurements. The thick blue line presents the model result, defined as $Q_f$, from the right-hand side of 
equation~\ref{govern_n2}. The uncertainty in the model results is estimated from the relative error as given by

\begin{eqnarray}
E = \Big| \frac{Q_m - Q_f}{Q_m} \Big|
\label{fittingerr}
\end{eqnarray}

The relative error is indicated by the connected triangles in red and plotted on right y-axis in the Figure~\ref{errfit}. In the plot, the value of $R$, $L$ and $D$  is kept in unit of \textit{R}$_\odot$ and value of $\gamma$=5/3 is set. The model result matches very well with the measurements. The relative errors of all the points are below 1\% when the CME is in COR2 FOV. We mention here that we have also attempted to fit the model to the measurements in COR-1 FOV and found that the  relative error reached more than 10 \% indicating a poor fitting. This is another reason, in addition to violation of self-similar expansion in 
COR1 FOV, to avoid inclusion of the analysis when the CME is in COR1 FOV. It has also been pointed out in Paper I that the neglect of gravitation force and the equivalent fictitious force due to the use of a non-inertial reference frame of the Sun may cause a lager error in FRIS model at the near-Sun region. Therefore, in the next section we present the internal properties of the CME only in the COR2 FOV.

\textbf{\subsection{Thermodynamic process} \label{intparam_n1}} 

\begin{figure}
\begin{center}
\includegraphics[angle=0,scale=0.60]{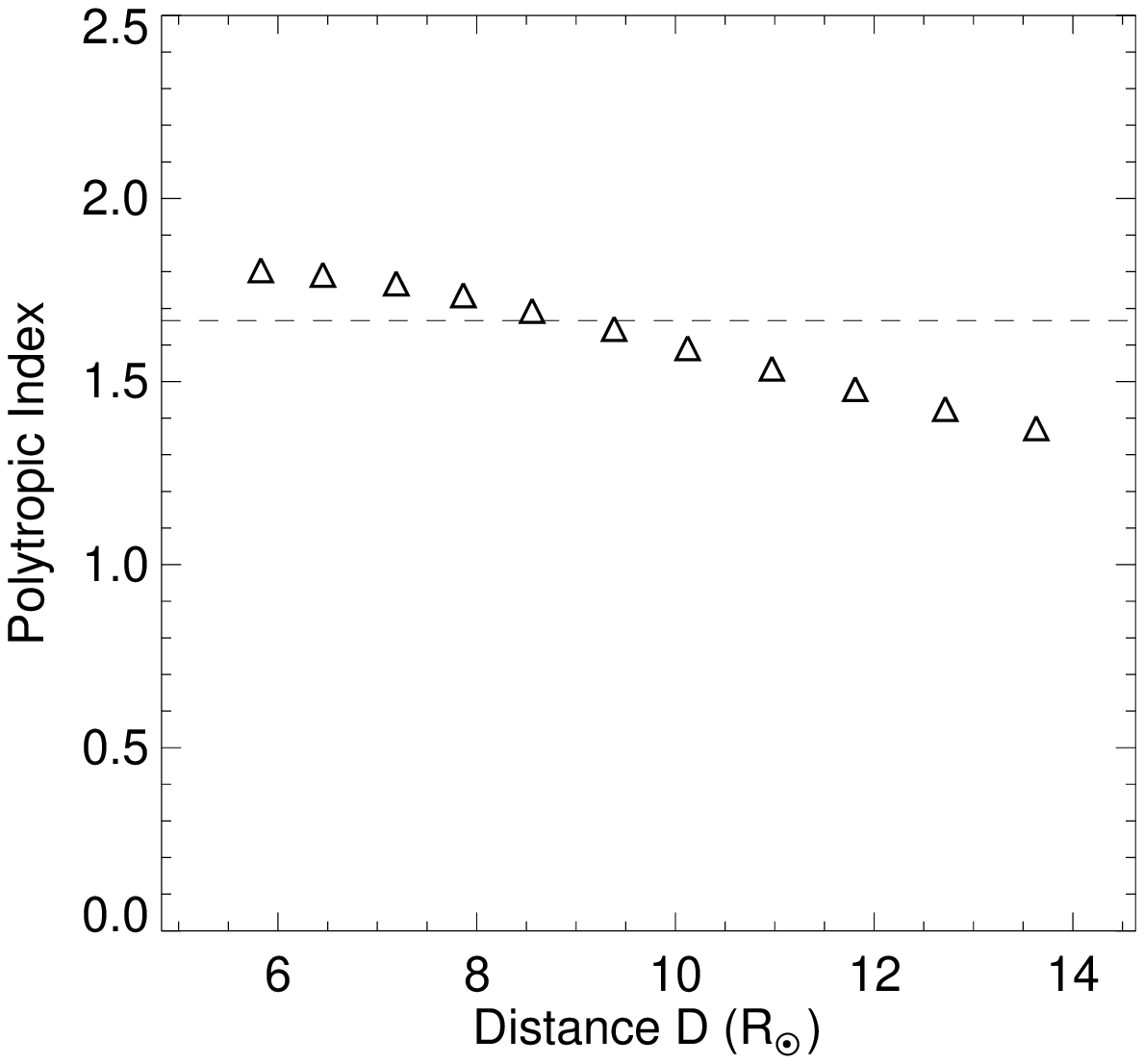}
\hspace{0.6cm}
\includegraphics[angle=0,scale=0.60]{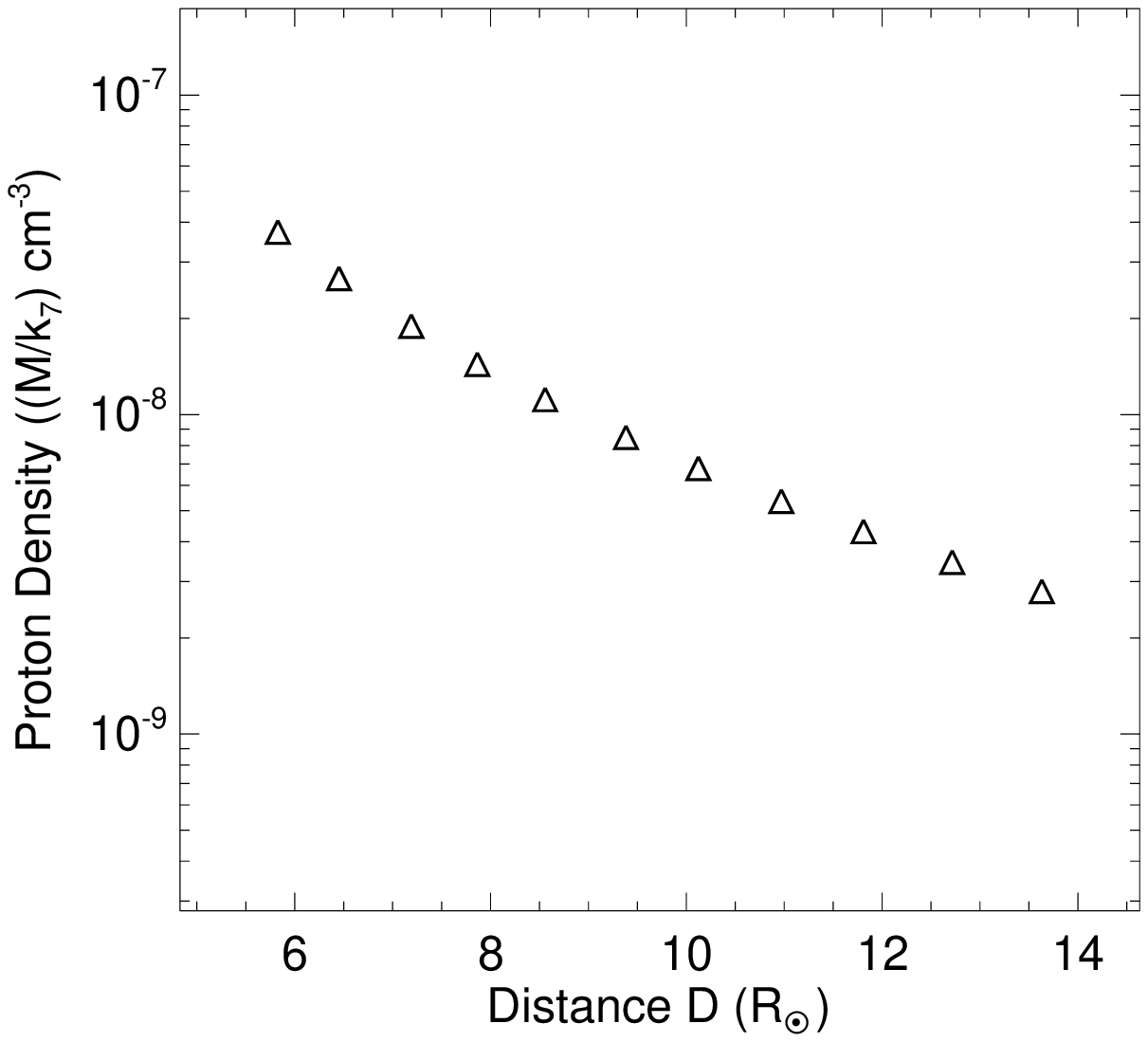}  \\
\vspace{0.2cm}
\includegraphics[angle=0,scale=0.60]{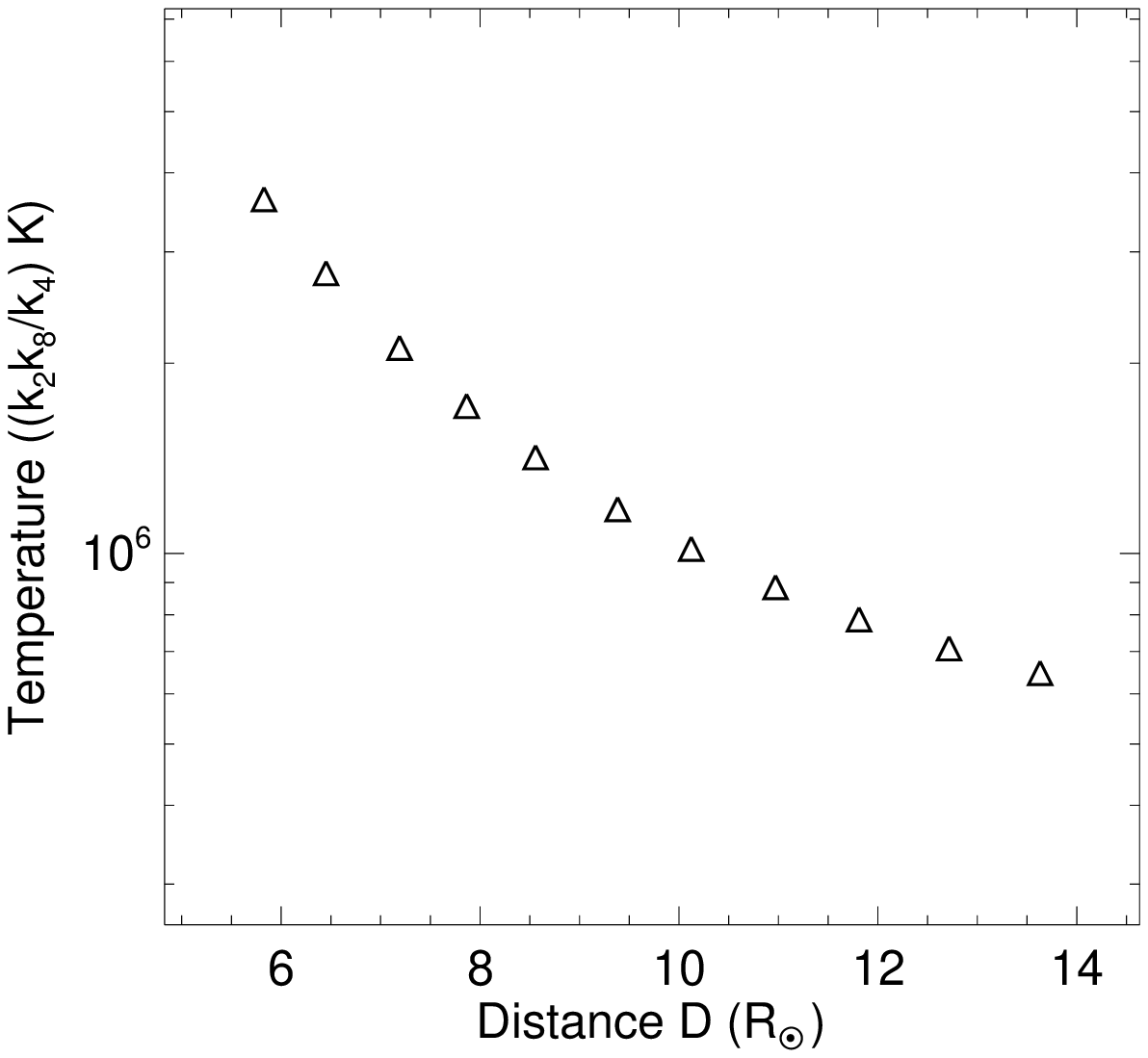}
\hspace{0.6cm}
\includegraphics[angle=0,scale=0.60]{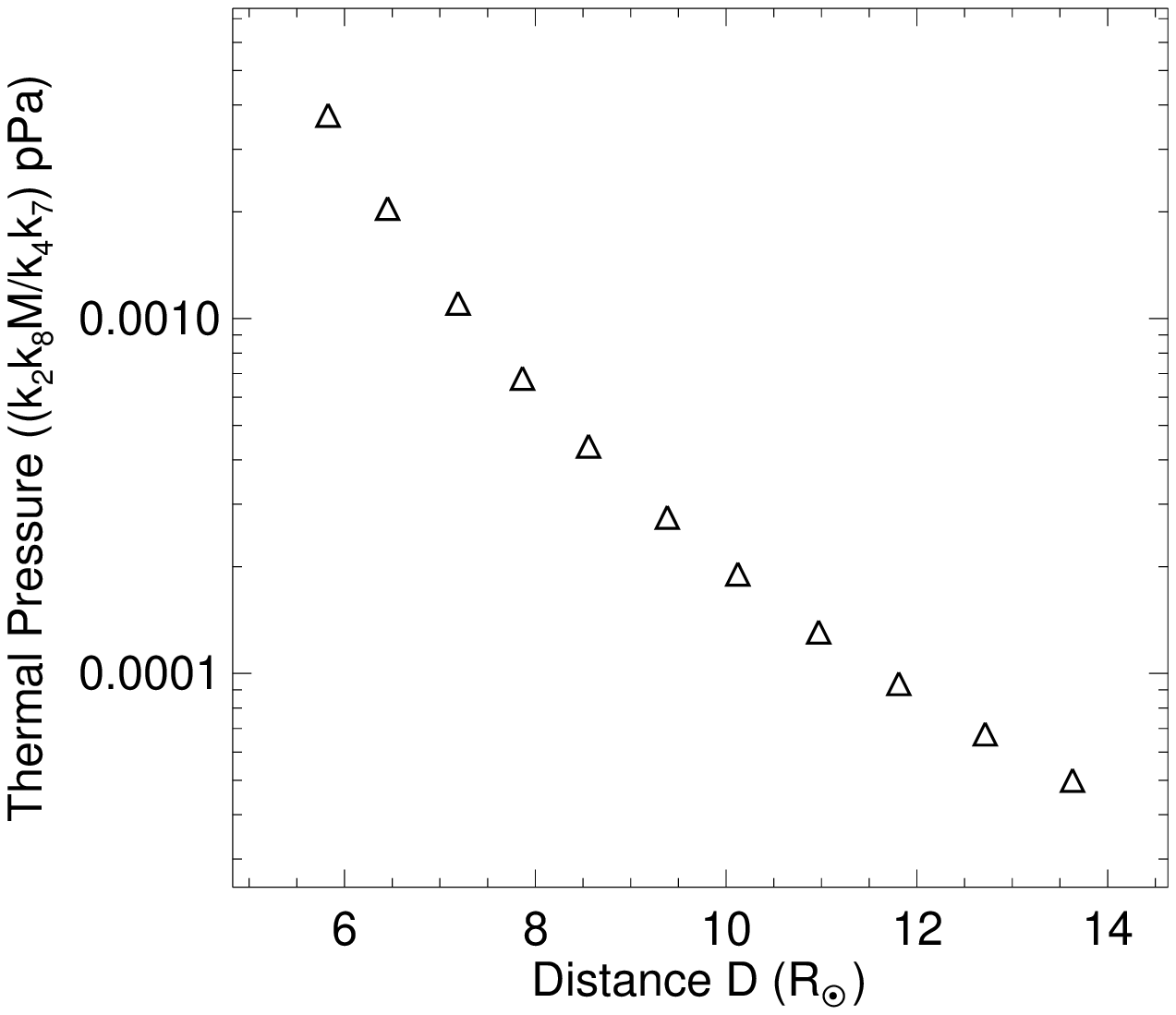}
\caption{The variation of the polytropic index ($\Gamma$), average proton number density ($\overline{n_{p}}$), Temperature ($\overline{T}$) and thermal pressure ($\overline{p}$) of the CME with heliocentric distance of its center ($D$) is shown in the top-left, top-right, bottom-left and bottom-right panel, respectively.}
\label{gamdentemthp}
\end{center}
\end{figure}

\textbf{\subsubsection{Polytropic index, density, temperature and thermal pressure} } 
We used the equation~\ref{dens_nn2}, ~\ref{thpres}, ~\ref{temp} and  ~\ref{polytropic_f} to estimate average proton number density ($\overline{n}_p$), thermal pressure ($\overline{p}$), temperature ($\overline{T}$) and polytropic index ($\Gamma$). The estimated parameters are shown in Figure~\ref{gamdentemthp}. It is 
noted that except for $\Gamma$, other parameters are given in a relative value because of the unknown factors. From the top-left panel of the figure, it is found that $\Gamma$ is decreasing as the CME is evolving through the corona. Near the Sun when the center of the CME is at a heliocentric distance of $D$ = 5.8 \textit{R}$_\odot$, the value of polytropic index is about 1.8. Its value is decreasing slowly and reaches to around 1.66 at 9 \textit{R}$_\odot$. The value of $\Gamma$ $>$ 1.66 means that CME is releasing the heat into surrounding. After $D$=9 \textit{R}$_\odot$,  the $\Gamma$ shows a faster decrease and approached down to about 1.37 meaning that the CME is absorbing the heat. It is also to be noted that although the propagation speed of the CME always increased, the expansion speed first decreased up to around 9 \textit{R}$_\odot$ and then begin increasing (Figure~\ref{kinem}).

CMEs are usually hot plasma structures originating from active regions mostly through magnetic field reconnection. Thus , initially, CMEs have a temperature larger than ambient solar wind \citep{Antonucci1997,Ciaravella2000}, and deliver the heat to the surrounding. The expansion of the CME leads to some work done by it and causes the temperature decreasing (as shown in bottom-left panel of Figure~\ref{gamdentemthp}). According to our model results, before $D$ = 9 \textit{R}$_\odot$, the expansion has not been sufficient, and therefore the value of $\Gamma$ is greater than 1.66.

The continuous expanding propagation of CME leads to a decrease of its density (top-right panel of Figure~\ref{gamdentemthp}) which in addition to decrease in temperature (bottom-left panel of the figure) result in a decrease in thermal pressure (bottom-right panel of the figure) inside the CME. At the beginning, a decrease in the proton density, temperature and thermal pressure is much faster and then becomes slower tending towards its asymptotic values. The exact physical mechanisms responsible for transfer of heat as energy need to be explored further. It is possible that heat is conducted from the lower corona to the CME as the looped structure of a CME is considered to be always attached with the Sun \citep{Gosling1987a,Larson1997}. As the temperature of the ambient solar wind is higher than that in CME in the outer corona, the ambient medium can be an additional source to heat the CME. However, the cross-field diffusion of particles are limited than the motion parallel to magnetic field lines \citep{Matthaeus2003,Qin2006,Ruffolo2008}.

\textbf{\subsubsection{Heat and entropy} \label{intparam_n2}}

\begin{figure}
\begin{center}
\includegraphics[angle=0,scale=0.60]{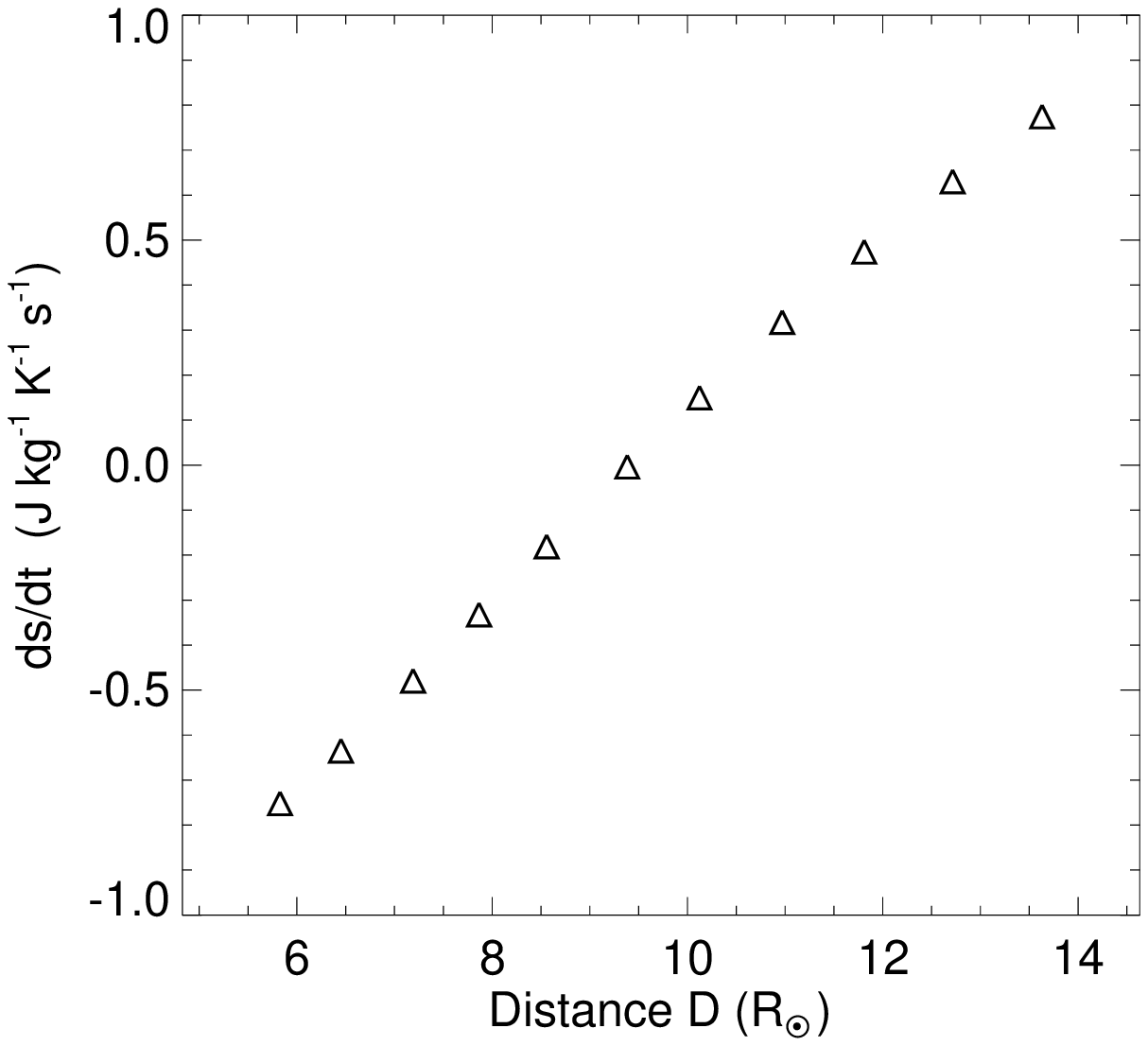}
\hspace{0.6cm}
\includegraphics[angle=0,scale=0.60]{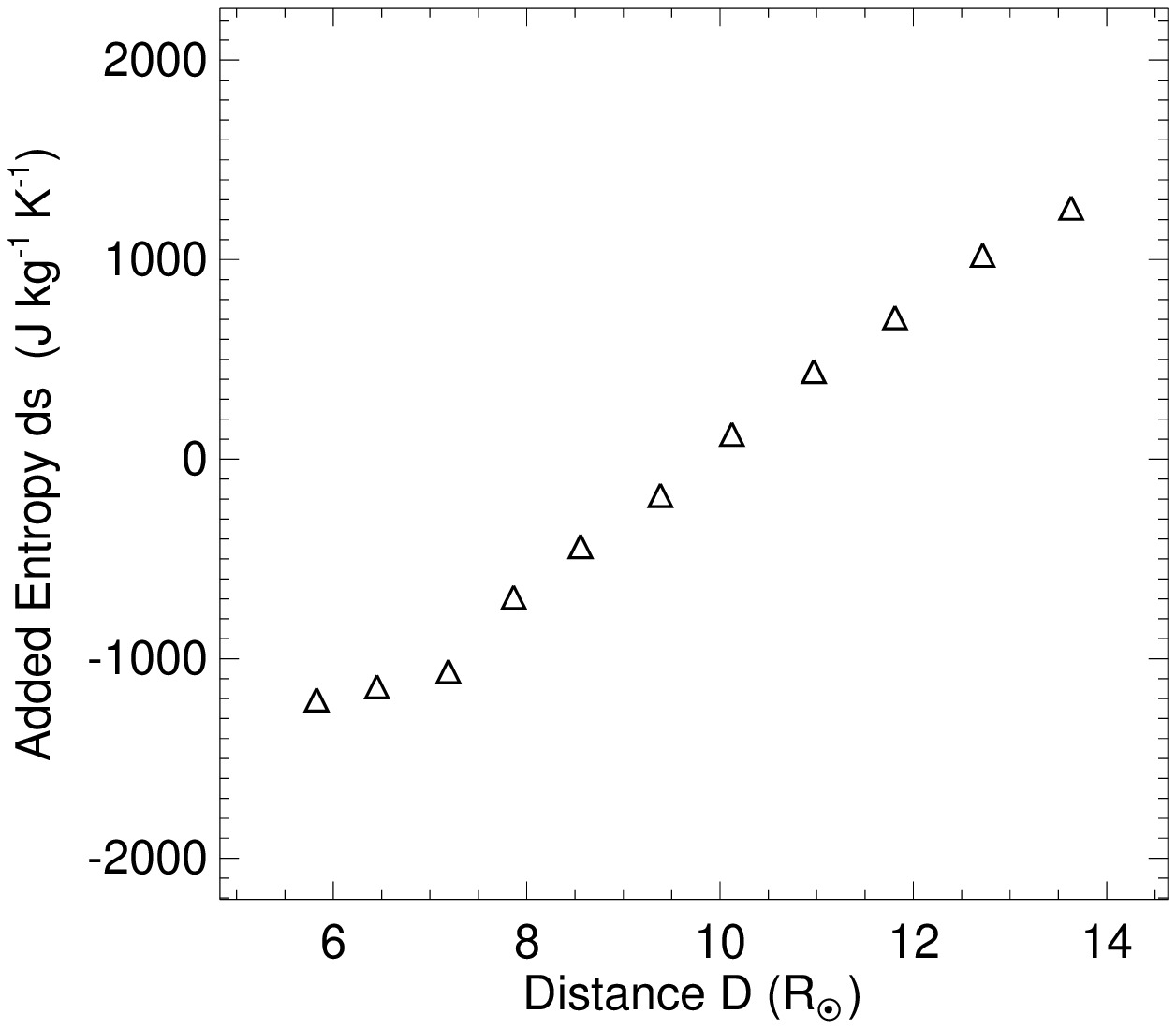}  \\
\vspace{0.2cm}
\includegraphics[angle=0,scale=0.60]{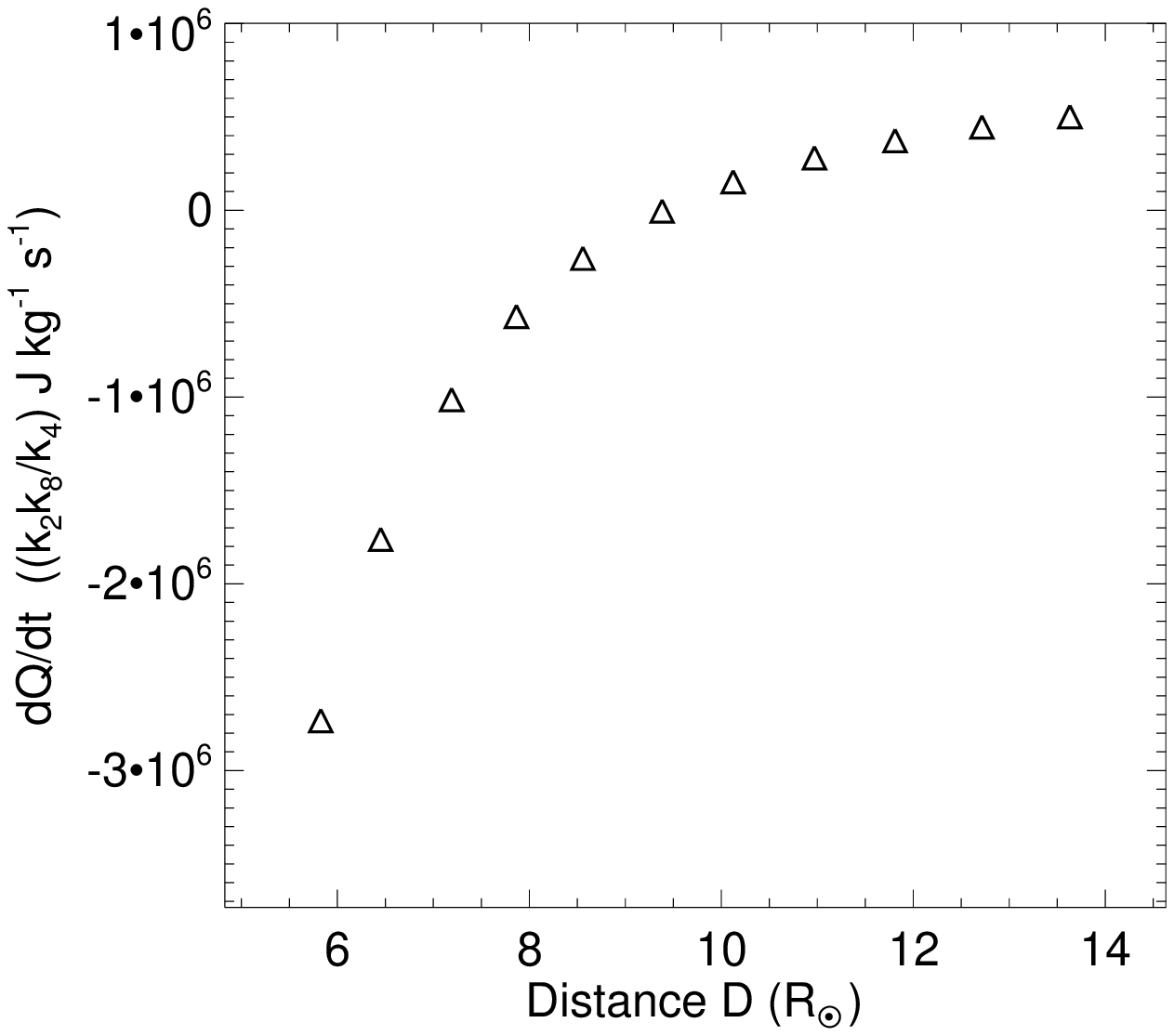}
\hspace{0.6cm}
\includegraphics[angle=0,scale=0.60]{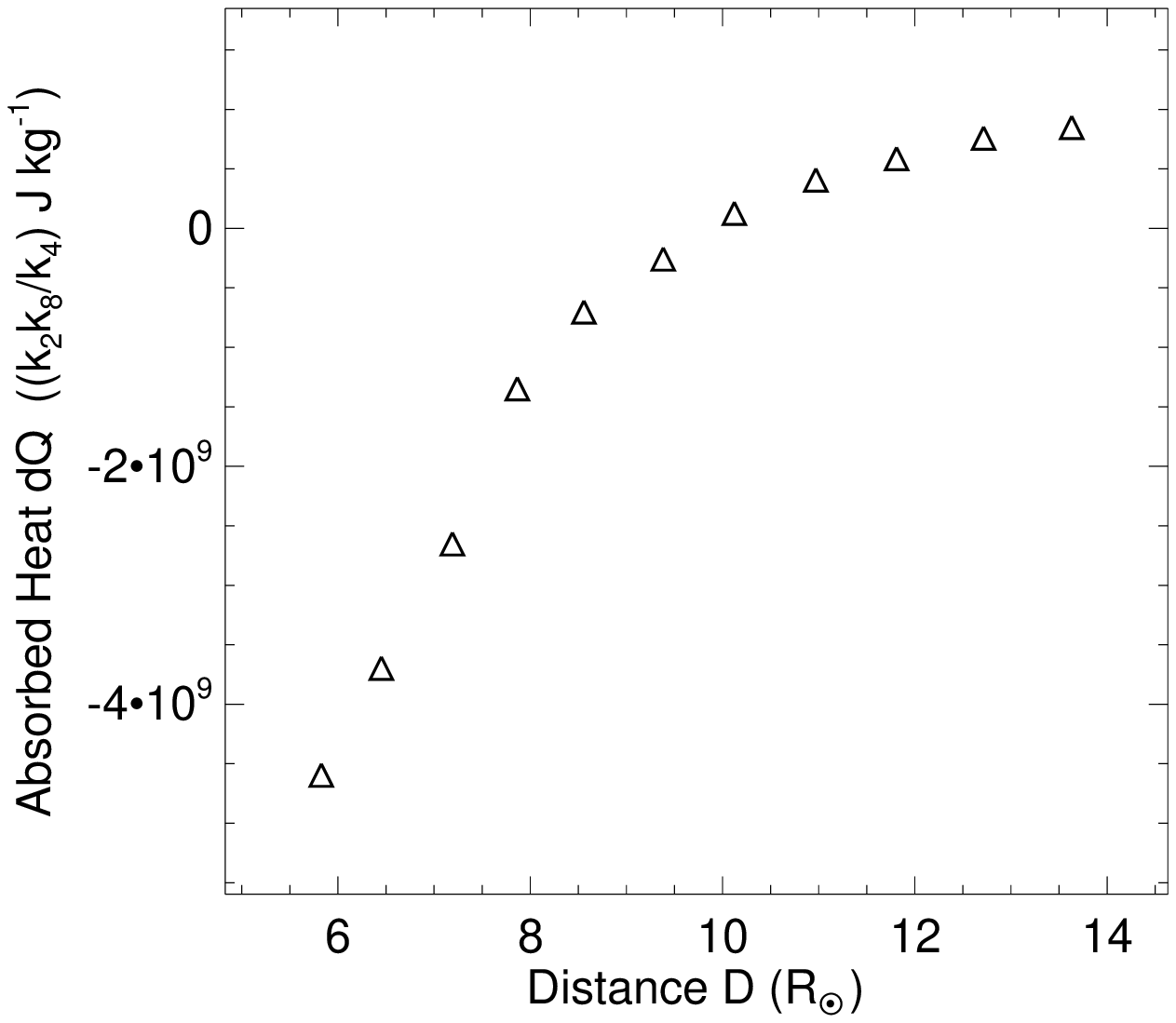}
\caption{The variation of the rate of change of entropy ($\frac{ds}{dt}$) of the CME, added entropy ($s$) to the CME, heating rate ($\overline{\kappa}$=$\frac{dQ}{dt}$) of the CME and absorbed heat ($Q$) by the CME with heliocentric distance of its center ($D$) is shown in the top-left, top-right, bottom-left and bottom-right panel, respectively.}
\label{dentdheat}
\end{center}
\end{figure}

The rate of change of entropy (i.e, $\frac{ds}{dt}$ = $ \frac {\overline{\kappa}}{\overline{T}}$) per unit mass and average heating rate (i.e., $\overline{\kappa}$ = $\frac{dQ}{dt}$) per unit mass is estimated using the equation~\ref{rtofentropy} and ~\ref{heatingrate}. The estimated value of 
$\frac{ds}{dt}$ is shown in the top-left panel of Figure~\ref{dentdheat}. The value of  $\frac{ds}{dt}$ derived by the model is absolute, unlike to other parameters 
(e.g., density, temperature, heat etc.) of which only the relative values are known. The plot of $\frac{ds}{dt}$ is very useful as the heating rate per mass of the CME plasma can be read from it if the real temperature was known. The value of $\frac{ds}{dt}$ was about - 0.75 J kg$^{-1}$ K$^{-1}$ s$^{-1}$ at the beginning and reached zero at about $D$ = 9 \textit{R}$_\odot$ and continuously increased to about 0.8 J kg$^{-1}$ K$^{-1}$ s$^{-1}$ at $D$ = 13.6 \textit{R}$_\odot$. It implies that the rate of loss of entropy is getting smaller as CME moved out and then after $D$ = 9 \textit{R}$_\odot$ the CME started to gain entropy with increasing rate. It is also clear from the top-right panel of the figure that the added entropy ($ds$) is negative in the beginning, i.e., release of entropy is taking place from the CME. Then the $ds$ increased towards positive value implying a transfer of entropy into the CME system.

The bottom-left panel of Figure~\ref{dentdheat} shows that the value of heating rate $\overline{\kappa}$=$\frac{dQ}{dt}$ was negative at the beginning and increased towards positive value with crossing zero at $D$=9 \textit{R}$_\odot$. The negative value of $\overline{\kappa}$ implies that CME is releasing the heat while its positive value implies that CME is absorbing the heat. The absorbed heat ($dQ$) by the CME is elegantly shown in bottom-right panel of the figure. The positive values of $dQ$ imply that thermal energy is being transferred from somewhere into the CME while its negative values mean that thermal energy is being released out from the CME. We note that the CME behaves like an adiabatic closed system at $D$=9 \textit{R}$_\odot$ where change in entropy is zero, and this suggests that no entropy is generated within the system boundaries.

\begin{figure}
\begin{center}
\includegraphics[angle=0,scale=0.60]{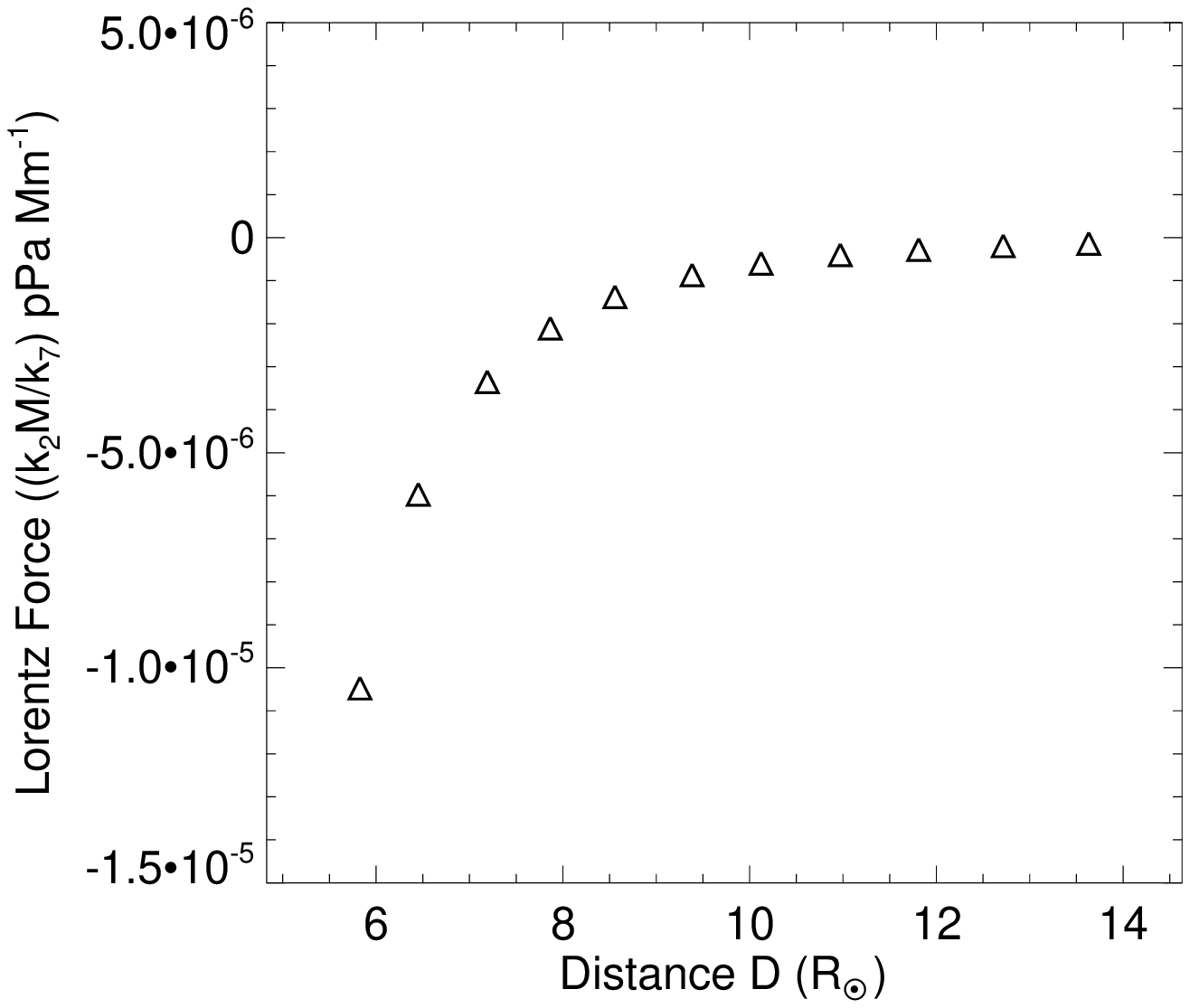}
\hspace{0.6cm}
\includegraphics[angle=0,scale=0.60]{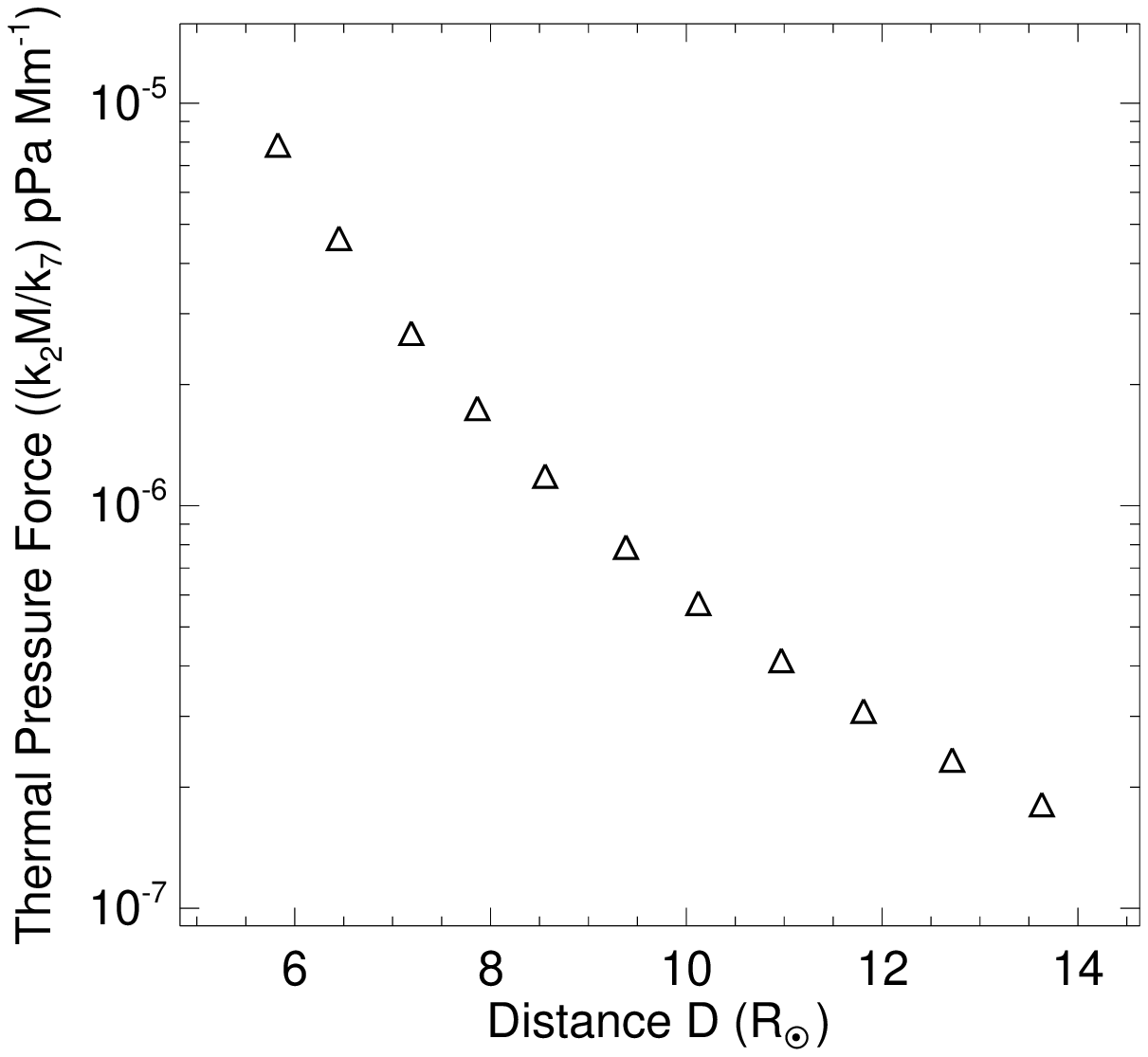}  \\
\vspace{0.2cm}
\includegraphics[angle=0,scale=0.60]{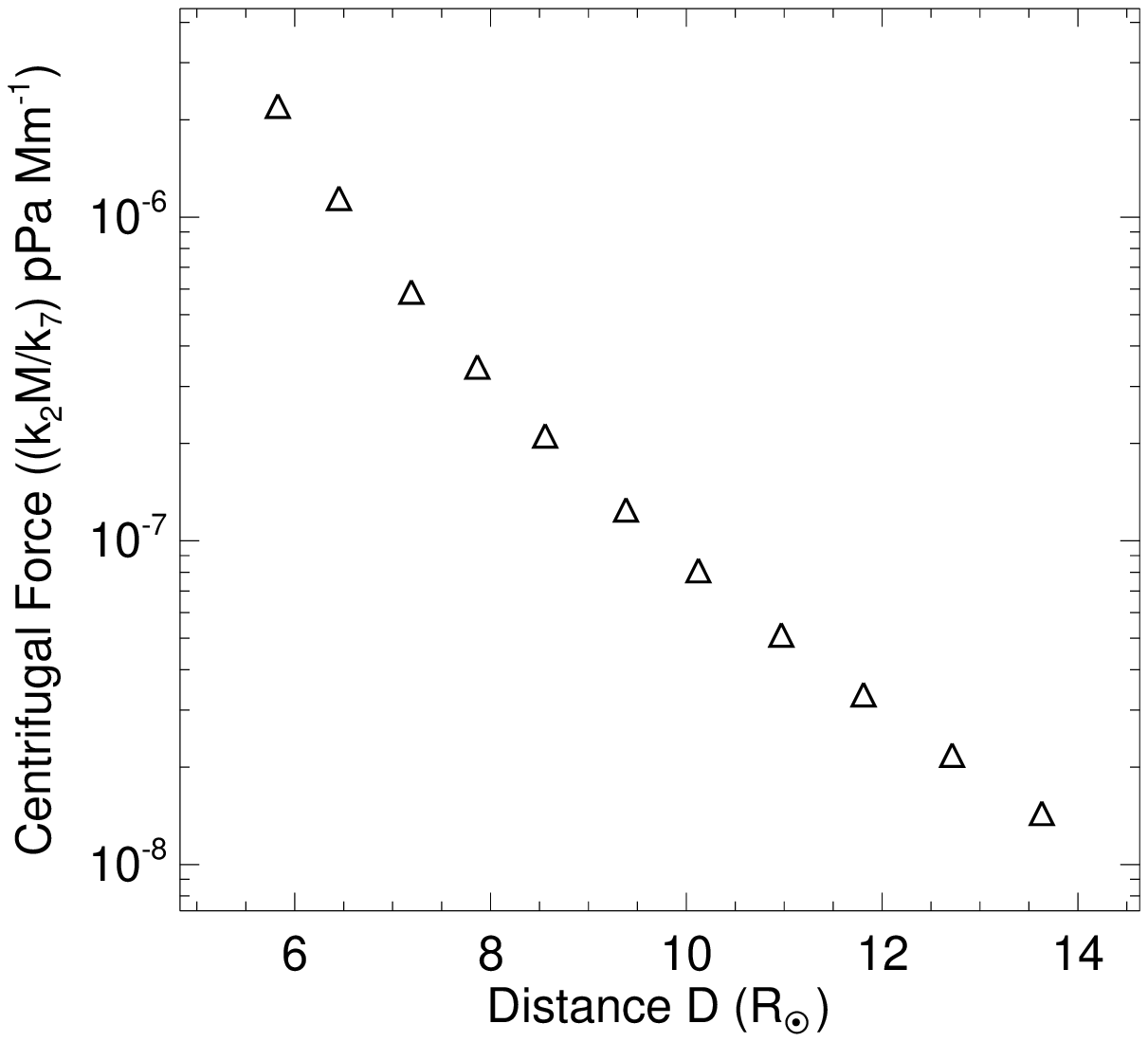}
\hspace{0.5cm}
\includegraphics[angle=0,scale=0.60]{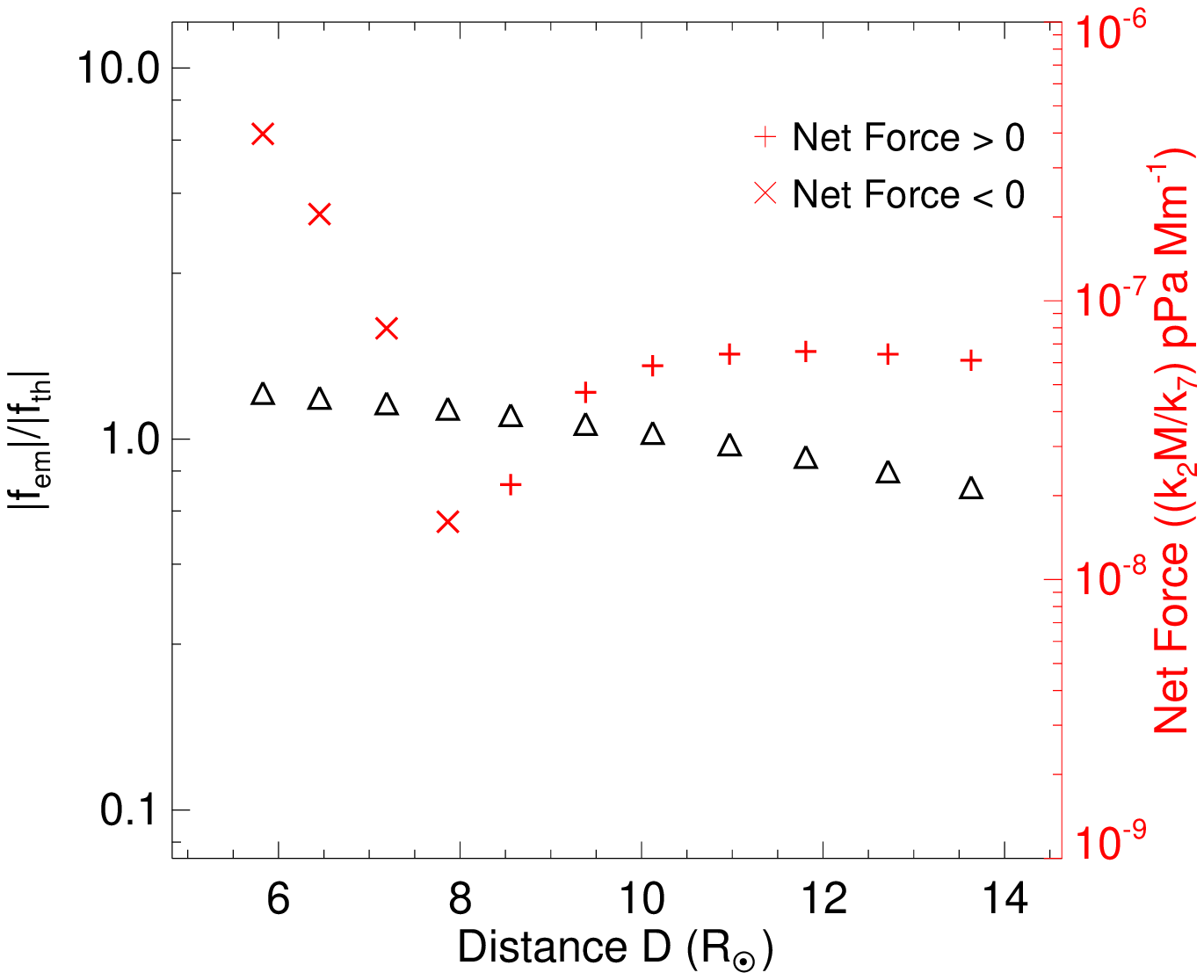}
\caption{The variation of different forces acting on the CME with heliocentric distance of its center ($D$) is shown. The average 
Lorentz force ($\overline{f}_{em}$), thermal pressure force ($\overline{f}_{th}$), and centrifugal force  ($\overline{f}_{p}$) are shown in top-left, top-right and bottom-left panel, respectively. The ratio of absolute values of Lorentz and thermal force (on left Y-axis) as well as the net force (on right Y-axis) acting inside the CME ($\overline{f}$) are shown in the bottom-right panel.}
\label{lorthrpolnet}
\end{center}
\end{figure}

\textbf{\subsection{Dynamic process} \label{intparam_n3}} 
\textbf{\subsubsection{Lorentz, thermal and centrifugal force}} 
The dynamic processes that could be deduced by our FRIS model are the changes of the average Lorentz force ($\overline{f}_{em}$), thermal pressure force 
($\overline{f}_{th}$) and the centrifugal force ($\overline{f}_{p}$) as the CME propagated away from the Sun and these are shown in Figure~\ref{lorthrpolnet}. 
From the top panel of the figure, it is noted that $\overline{f}_{em}$ have negative values whereas there are positive values for the $\overline{f}_{th}$. Thus, the directions of the two forces are opposite which indicate that $\overline{f}_{em}$ is acting towards the center of the CME and $\overline{f}_{th}$ is acting away from the center of the CME. The force directions suggest that (1) the $\overline{f}_{em}$ prevented the CME from free expansion, and also maintained a possible
poloidal motion of the CME plasma, and (2) the $\overline{f}_{th}$ was the internal cause of the CME expansion. The bottom-left panel of the figure shows that the the $\overline{f}_{p}$ drops quickly with around 85\% decrease before the center of the CME reaches 8 \textit{R}$_\odot$ and the decrease continues further asymptotically reaching zero.

The expansion acceleration of the flux rope plasma is driven by Lorentz force + thermal pressure gradient force - centripetal force, which is what Equation~\ref{lorentz_n1} describes (i.e., the net force $\overline{f}$ = $\overline{f}_{em} + \overline{f}_{th} + \overline{f}_{p}$). The centripetal force is a part of the sum of the Lorentz force and thermal pressure gradient force. Thus, the poloidal motion has nothing to do with the expansion acceleration. The net force ($\overline{f}$) inside the CME and the ratio of the average Lorentz to thermal forces ($\frac{\overline{f}_{em}}{\overline{f}_{th}}$) is shown in the bottom-right panel of the Figure~\ref{lorthrpolnet}. It is evident that the absolute values of both the $\overline{f}_{em}$ and $\overline{f}_{th}$ forces are getting very close to each other as the CME moves outward. The ratio of the two forces indicates that the $\overline{f}_{em}$ is stronger than the $\overline{f}_{th}$ since the beginning to 10 \textit{R}$_\odot$ where they become almost the same in the magnitude. After this distance, $\overline{f}_{th}$ becomes larger in the magnitude than the $\overline{f}_{em}$. Thus, the ratio of their absolute values range between 1.33 at the beginning and 0.74 at the last data point. The net force inside the CME is negative at the first four data points and this is expected as the expansion acceleration is less than zero  (i.e., negative) corresponding to these points as shown in Figure~\ref{kinem}. Starting from $D$=8.5 \textit{R}$_\odot$ to 13.6 \textit{R}$_\odot$, the net force is positive corresponding to the positive expansion acceleration of the CME.

We note that there are three data points between around 8.5 and 10 \textit{R}$_\odot$ where the expansion acceleration is positive despite having larger inward $\overline{f}_{em}$ than $\overline{f}_{th}$. Thus, the poloidal motion in the CME tends to contribute in the expansion and its presence suggests that flux-rope carry some angular momentum. Such an angular momentum may be generated locally through the interaction with ambient solar wind through viscosity \citep{Wang2015,Zhao2017,Zhao2017a}. Although there is no direct evidence about the plasma rotation inside a CME near the Sun, the rotation signature is indeed found in in-situ measurements of magnetic clouds at 1 AU \citep{Wang2015,Wang2016,Zhao2017a,Zhao2017}. Based on a statistical study, it is found in \citet{Wang2015} that the poloidal motion is not as significant as expanding motion generally. They noted that the median value of the poloidal speed was about 10 km s$^{-1}$ at 1 AU, while for a few cases the modeled poloidal speed was two to four times larger than the modeled expansion speed. Therefore, it was suggested that poloidal motion may also be recognized in observations. Further verification from the observations is needed to confirm the presence of such poloidal motion inside a CME near the Sun.

\textbf{\section{Summary and Discussion} \label{Resdis}}

\begin{table}
\caption{List of the derived internal thermodynamic parameters, constants and coefficients from FRIS model}
\begin{center}
\hrule
\vspace{0.1cm}
\hrule 
{\small
 \begin{tabular}{cccc} 
\multicolumn{4}{c}{\textit{Internal thermodynamic parameters derived from the model}} \\  \hline 
Quantities & Factors & Values & SI units \\   \hline
 Lorentz force ($\overline{f}_{em}$)            &  $\frac{k_2 M}{k_7}$           & $c_2 R^{-5} + c_3 L^{-2} R^{-3}$              & Pa m$^{-1}$  \\
Thermal pressure force ($\overline{f}_{th}$)   &  $\frac{k_2 M}{k_7}$           & $\lambda L^{-\gamma} R^{-\gamma-1}$           & Pa m$^{-1}$ \\
Centrifugal force ($\overline{f}_{p}$)   &  $\frac{k_2 M}{k_7}$           & $c_1 R^{-5} L^{-1}$                           & Pa m$^{-1}$ \\
Proton number Density ($\overline{n}_p$)      &  $\frac{M}{k_7}$               & $\frac{1}{\pi m_{p}}$ $(L R^2)$$^{-1}$     & kg m$^{-3}$ \\
Thermal pressure ($\overline{p}$)              &  $\frac{k_2 k_8 M}{k_4 k_7}$   & $\lambda (L R^2)^{-\gamma}$                   & Pa          \\
Temperature ($\overline{T}$)               &  $\frac{k_2 k_8}{k_4}$   & $\frac{\pi \sigma}{\gamma-1} \lambda (LR^2)^{1-\gamma}$ & K \\
Changing rate of entropy ($\frac{ds}{dt}$) &                          & $\frac{1}{\sigma \lambda} \frac{d\lambda}{dt}$          & J K$^{-1}$ kg$^{-1}$ s$^{-1}$  \\
Heating rate ($\overline{\kappa}$)     & $\frac{k_2 k_8}{k_4}$    & $\frac{\pi}{\gamma-1}(LR^2)^{1-\gamma} \frac{d\lambda}{dt}$ & J kg$^{-1}$ s$^{-1}$ \\
Thermal energy ($E_i$)                 & $\frac{k_2 k_8 M}{k_4}$  & $\frac{\pi }{\gamma-1} \lambda (L R^2)^{1-\gamma}$         & J \\
Magnetic energy ($E_m$)                &                          & $E_{m1}$ + $E_{m2}$                                             & J \\
$E_{m1}$                                 & $\frac{k_9}{k_7}$        & $\frac{\pi}{\mu_0} L^{-1}$                                 & J \\
$E_{m2}$                                 & $k_7 k_{10}$             & $\frac{\pi}{\mu_0} L R^{-2}$                               & J \\
Polytropic index ($\Gamma$)            &         & $\gamma+\frac {\ln \frac{\lambda(t)}{\lambda(t+\Delta t)}} {\ln \left\{\frac{L (t+\Delta t)}{L (t)} 
                                                  \Big[ \frac{R(t+\Delta t)}{R (t)} \Big]^2 \right\}}$                          &    \\				
\end{tabular}

\hrule 
\begin{tabular}{cc} 
\multicolumn{2}{c}{\textit{All the constants ($k_{1-12}$) introduced in the model}} \\  \hline
Constants &   Interpretations              \\  \hline
$k_1$            & Scale the magnitude of the poloidal motion \\  
$k_{2-6, 8-10}$  & Integrals of distributions of density, poloidal speed and magnetic vector potential  \\ 
$k_7$            & Ratio of the length of the flux rope $l$ to the distance $L$ \\
$k_{11}$         & Coefficient of equivalent conductivity \\
$k_{12}$         & Aspect ratio, i.e., the ratio of the radius of the flux rope $R$ to the distance $L$ \\ 
\end{tabular}

\hrule
\begin{tabular}{lc}
\multicolumn{2}{c}{\textit{All the coefficients ($c_{0-5}$) introduced in the model}} \\  \hline 
Coefficients   & Expressions  \\ \hline
$c_0$   &  $\frac{k_4 M^{\gamma-1}}{k_2 k_7^{\gamma-1}}$  \\ 
\noalign{\vskip 0.1cm}
$c_1$   &  $\frac{k_1^2 k_3 L_A^2}{k_2 M^2} \ge 0$    \\
\noalign{\vskip 0.1cm}
$c_2$   &  $\frac{- k_5 k_7}{\mu_0 k_2 M}$     \\ 
\noalign{\vskip 0.1cm}
$c_3$   & $\frac{- k_6}{\mu_0 k_2 k_7 M}  \le 0$   \\ 
\noalign{\vskip 0.1cm}
$c_4$   & $\frac{k_2 k_8 M}{(\gamma-1) k_4 k_7 k_{11} T_a}$  \\
\noalign{\vskip 0.1cm}
$c_5$   & $\frac{\pi \sigma k_2 k_8}{(\gamma-1) k_4 T_a}$ \\ 
\noalign{\vskip 0.1cm}
\end{tabular}}
\hrule
\vspace{0.1cm}
\hrule 
\label{summ}
\end{center}
\tablecomments{\scriptsize Top panel: Quantity=Factor $\times$ Value. The factors are unknown constants that can not be inferred from FRIS model. The distance ($L$) of the center of flux rope CME from the solar surface and its radius ($R$) can be derived from the observations. The variable $\lambda$ is given by 
equation~\ref{govern_n2}. The $\gamma$ is the heat capacity ratio (adiabatic index) and $\mu_{0}$ is the magnetic permeability of free space. The  
$\sigma$=$\frac{(\gamma-1)m_p}{2k}$, where $m_p$ is the proton mass and $k$ is the Boltzmann constant, $M$ is the total mass of a CME. Middle panel: Among the constants, $k_{2,7,8,11} > 0$ and $k_{3,6,9,10} \ge 0$.  Bottom panel: The coefficients ($c_{0-5}$) are also constants and can be derived from the model. $L_A$ is the total angular momentum of a flux rope CME and $T_a$ is the equivalent temperature of the ambient solar wind around the CME base. \normalsize}
\end{table}

In the present study, we have improved the analytical flux rope internal state (FRIS) model for investigating the evolution of the internal thermodynamic state of a CME. The model is constrained by the observed propagation and expansion behavior of a CME. The internal thermodynamic parameters derived from the model and the details of the constants introduced in the model are summarized in the Table~\ref{summ}. We have applied the FRIS model to the CME of 2008 December 12 and determined the evolution of its internal properties. We find that the polytropic index of the CME plasma decreased from initially 1.8 to 1.66 slowly and further quickly decreased down to around 1.35. It suggests that initially there be heat released out from the CME before reaching to an adiabatic state and then a continuous injection of heat into the CME plasma. The value of polytropic index obtained in our study is not in good agreement with that obtained in 
\citet{Liu2005,Liu2006} using combined surveys of ICMEs in in situ observations. However, their estimated polytropic index of 1.1 to 1.3 was for the distance between 
0.3 and 20 AU while the CME we have studied was within 15 solar radii from the Sun.

We also find that the Lorentz force is directed inward while the thermal pressure force and centrifugal force is directed outward from the center of 
the CME. All these three forces decreased as the CME propagated away from the Sun. The time-variation in resultant direction of the net force is consistent with the expansion acceleration which reveals that the thermal pressure force is the internal driver of the CME expansion, whereas the Lorentz force prevented the CME from expansion. We emphasize that, in general, the direction of Lorentz force may be both inward or outward, i.e., the Lorentz force can cause both expansion and contraction of the flux-rope. As it is obvious from equation~\ref{lorentz_n2} that the constant $k_6$ is always larger than or equal to zero, while the sign of the constant $k_5$ is determined by the $B_z^2 (R) - B_z^2 (0)$. It implies that $J_\phi$ $\times$ $B_z$ could contribute in contraction or expansion depending on the distribution of $B_z$ in the cross-section of the flux rope. Also, it can be noted from equation~\ref{emforce} that the direction of Lorentz force depends on the unknowns $c_2$ and $c_3$. The coefficient $c_3$ is always less or equal to zero while the sign of the coefficient $c_2$ depends on the constant $k_5$. For the CME under study, on fitting equation~\ref{govern_n1n2}, the value of $c_2$ was estimated to be less than zero. This implies that the sign of 
$k_5$ was positive. Therefore, the direction of Lorentz force was inward which prevented the CME from expansion. Based on the obtained results, we note that the negative expansion acceleration, Lorentz force dominating the thermal force and release of heat happen together. Moreover, we find that among the three forces, the centrifugal force decreased with fastest rate and Lorentz force decreased slightly faster than the thermal pressure force.  We note that even a small difference between the Lorentz and thermal forces can drive the expansion acceleration of the CME at the order of 1 m s$^{-2}$.

As a consequence of the release and absorption of heat by the CME, a decrease in the entropy during initial phase and an increase in the entropy at latter phase of the CME propagation is found. Our analysis find that rate of heat release and entropy loss of the CME is slowing down with time, and gradually turning into increasing rate of heat absorption and entropy gain. Thus, the CME studied here is found to go through an isentropic (adiabatic and reversible, no heat transfer) point 
around 9 \textit{R}$_\odot$ after which the expansion acceleration of the CME is positive. Since, heat is disorganized form of energy, the direction of entropy transfer is the same of the direction of heat transfer.

This CME launched with hot temperature from lower corona of the Sun through a rapid magnetic energy release process and then cooled down during expansion. The heat release out of the CME in its initial propagation phase before around 9 \textit{R}$_\odot$ is probably due to that the CME structure was not expanded sufficiently and therefore still hotter enough than ambient coronal medium. However, the CME is found to be heated in the latter phase of its propagation. The heating could be possibly due to the conduction of heat from the solar surface, solar wind or due to dissipation of magnetic energy into thermal energy. \citet{Kumar1996} suggested that the heating may result from the local magnetic dissipation as they have shown that around 58\% of magnetic energy lost in the expansion is available for the heating.

In our study, we attempted to apply FRIS model to the complete measurements from COR1 and COR2 observations. However, we faced difficulty in fitting the governing 
equation~\ref{govern_n2} for the CME measurements in COR1 FOV, for which the model was not reliable. We think that the rising and declining phase of the CME acceleration in COR1 FOV and then a gap of measurements between COR1 and COR2 FOVs, could be the reasons for the unreliability. We also tried to separate the observations of acceleration and deceleration phase and then fit the each phase with the model. This also could not improve the fitting as the number of data points were not sufficient enough. Based on these trials, we decided not to include the COR1 measurements in the present study. This is logical as the CME may not follow the self-similar expansion, which is one of the basic assumptions of the FRIS model, during its impulsive acceleration phase near the Sun. Moreover, the gravitational forces and the equivalent fictitious force in a non-inertial frame are not considered in FRIS model. Although these forces are very small for most CMEs beyond outer corona, however they may significantly distort the model results for some CMEs which have slow expansion acceleration in the inner corona near the Sun.

Further, there have been studies on the solar wind stretching effect caused by the divergent radial expansion of solar wind flow, which is of a kinematic effect, on the expansion of CMEs \citep{Newkirk1981,Suess1988,Crooker1996,Riley2003,Riley2004}. Due to this effect, a flux rope CME in the solar wind does not remain in cylindrical shape but rather becomes a convex-outward, ``pancake'' shape. These studies confirm that the self-similar assumption is broken gradually as CME moves away from the Sun. Earlier studies have suggested that self-similar assumption is a valid approximation when the CME is nearly force-free and not too far away 
from the Sun \citep{Low1982,Burlaga1988,Chen1997,Demoulin2009,Shapakidze2010,Subramanian2014}. We expect that the FRIS model would be suitable to apply on the CME observations from couple of solar radii (i.e, COR2 FOV) to within tens of solar radii from the Sun. In another study, we plan to estimate the internal parameters of the CME beyond coronagraphic FOV using CME measurements from HI1 observations.

In the FRIS model, some effects of coupling between the ambient solar wind and the CMEs have been implicitly included. The drag interaction between the CMEs and the solar wind, causing momentum exchange between them, has effect on the propagation speed (i.e., $v_c$) of the CMEs \citep{Cargill2004,Manoharan2006,Vrsnak2010,Subramanian2012,Mishra2013,Liu2016}. Further, the pressure in ambient solar wind has effect of preventing the free expansion (i.e., $v_e$) of CMEs \citep{DalLago2003,Demoulin2009,Gulisano2010}. These effects of solar wind on a CME give different time-variation of $L$ and $R$, which can be measured from imaging data, and therefore are included in the model. Moreover, the distortion of the circular cross section of the flux rope is not considered mainly because the difficulty of the derivation of the model.

Following the discussion in Paper I, another limitation of the model is due to the assumption of an axisymmetric cylindrical flux rope, and thus the curvature of the axis of the flux rope is not considered in the model. The axial curvature causes the Lorentz force to have an additional component for driving the CME in the heliosphere. The neglect of the axial curvature generally would cause an underestimation of Lorentz force from the model. We emphasize that distortion of the circular cross-section of the flux rope would cause an overestimation of the its radius ($R$) and underestimation of its distance ($L$) from the solar surface. Therefore, it is expected that from our measurements the expansion speed is overestimated and the propagation speed is underestimated. Thus we find that in our study, the density, thermal pressure force and Lorentz force is underestimated than its actual values.  However, the extent of underestimation would be different at different distances from the Sun. We expect that errors in the derived internal thermodynamic parameters would not be significant at smaller distances, i.e., within COR2 FOV, where the distortion in the circular cross-section is not much.

The FRIS model described in this study is an improved version of the earlier developed flux rope model in Paper I. In Paper I, the authors have made a comparison of their model with \citet{Kumar1996}, and the differences described there would also be valid for the improved version of the model. It would be interesting to extrapolate the estimated internal thermodynamic properties of the CMEs to far away from the Sun up to around 1 AU. This can be done by extrapolating the observed kinematics using drag based model \citep{Vrsnak2013} of the CMEs or directly measuring the kinematics from imaging observations. The uncertainty from the model can further be examined based on the CME measurements from in situ observations. In a separate study, we would focus on the quantification of errors due to breakdown of some of the assumptions taken in the model. Further, it would be worth examining if other CMEs also show a similar evolution of their thermodynamic parameters as 2008 December 12 CME studied here.

\newpage

\textbf{\section*{Acknowledgments}}
We acknowledge the UK Solar System Data Center (UKSSDC) for providing the STEREO/COR2 data. Y.W. is supported by the National Natural Science Foundation of China 
(NSFC) grant Nos. 41574165, 41774178 and 41761134088. W.M. is supported by the NSFC grant No. 41750110481 and Chinese Academy of Sciences (CAS) President's International Fellowship Initiative (PIFI) grant No. 2015PE015.

%\bibliographystyle{aasjournal}
%\bibliography{wageesh_ref}

\end{document}